\documentclass{JINST}

\usepackage{rotating}
\usepackage{array}

\title{Assembly and Installation of the Daya Bay Antineutrino Detectors}

\author{H.R.~Band$^a$\thanks{Corresponding
author.}, 
R.L.~Brown$^b$,
R.~Carr$^c$,
X.C.~Chen$^d$,
X.H.~Chen$^e$,
J.J.~Cherwinka$^f$,
M.C.~Chu$^d$,
E.~Draeger$^g$,
D.A.~Dwyer$^{c,h}$,
W.R.~Edwards$^{h,i}$,
R.~Gill$^b$,
J.~Goett$^j$,
L.S.~Greenler$^f$,
W.Q.~Gu$^k$,
W.S.~He$^l$,
K.M.~Heeger$^a$,
Y.K.~Heng$^e$,
P.~Hinrichs$^a$,
T.H.~Ho$^k$,
M.~Hoff$^h$,
Y.B.~Hsiung$^l$,
Y.~Jin$^e$,
L.~Kang$^m$,
S.H.~Kettell$^b$,
M.~Kramer$^{h,i}$,
K.K.~Kwan$^d$,
M.W.~Kwok$^d$,
C.A.~Lewis$^a$,
G.S.~Li$^i$,
N.~Li$^h$,
S.F.~Li$^n$,
X.N.~Li$^e$,
C.J.~Lin$^h$,
B.R.~Littlejohn$^{a,o}$,
J.L.~Liu$^{c,k}$,
K.B.~Luk$^{h,i}$,
X.L.~Luo$^e$,
X.Y.~Ma$^e$,
M.C.~McFarlane$^a$,
R.D.~McKeown$^{c,n}$,
Y.~Nakajima$^h$,
J.P.~Ochoa-Ricoux$^h$,
A.~Pagac$^f$,
X.~Qian$^{b,c}$,
B.~Seilhan$^g$,
K.~Shih$^d$,
H.~Steiner$^{h,i}$,
X.~Tang$^e$,
H.~Themann$^b$,
K.V.~Tsang$^h$,
R.H.M.~Tsang$^c$,
S.~Virostek$^{h}$,
L.~Wang$^{e}$,
W.~Wang$^{a,n}$,
Z.M.~Wang$^e$,
D.M.~Webber$^a$,
Y.D.~Wei$^m$,
L.J.~Wen$^e$,
D.L.~Wenman$^f$,
J.~Wilhelmi$^j$,
M.~Wingert$^h$,
T.~Wise$^a$,
H.L.H.~Wong$^{h,i}$,
F.F.~Wu$^c$,
Q.~Xiao$^f$,
L.~Yang $^m$,
Z.J.~Zhang $^m$,
W.L.~Zhong $^{e,h}$,
~and H.L.~Zhuang$^e$\\
\llap{$^a$}Department of Physics, University of Wisconsin - Madison,\\
  1150 University Avenue, Madison, WI 53706, U.~S.~A.\\ 
 \llap{$^b$}Brookhaven~National~Laboratory, PO Box 5000, Upton, NY 11973,  U.~S.~A.\\
\llap{$^c$}Kellog Radiation Laboratory, California~Institute~of~Technology, Pasadena, CA 91125, U.~S.~A.\\
 \llap{$^d$}Department of Physics, Chinese University of Hong Kong, Shatin, Hong Kong, China\\
 \llap{$^e$}Institute of High Energy Physics, 19 Yuquan Road, Beijing, China\\
\llap{$^f$}Physical Sciences Laboratory, University of Wisconsin - Madison,\\
  3725 Schneider Drive, Stoughton, WI 53589, U.~S.~A\\
 \llap{$^g$}Department of Physics, Illinois~Institute~of~Technology, \\
 3101 South Dearborn St., Chicago, IL 60616, U.~S.~A.\\ 
 \llap{$^h$}Lawrence Berkeley National Laboratory, 1 Cyclotron Road, Berkeley, CA 94720, U.~S.~A.\\
 \llap{$^i$}Department of Physics, University of California, Berkeley, CA 94720, U.~S.~A.\\
  \llap{$j$}Department of Physics, Rensselaer~Polytechnic~Institute, \\
   110 Eight Street, Troy, NY 12180, U.~S.~A. \\ 
 \llap{$^k$}Department of Physics, Shanghai~Jiao~Tong~University, Shanghai, China\\  
 \llap{$^l$}Department~of~Physics, National~Taiwan~University, No. 1 Sec 4, Taipei, Taiwan\\
 \llap{$^m$}Dongguan Institute of Technology, Dongguan, Guangdong, China \\ 
 \llap{$^n$}Department of Physics, College of William and Mary, \\
  116 Jamestown Rd., Williamsburg, VA 23187, U.~S.~A. \\
 \llap{$^o$}Department of Physics, University~of~Cincinnati \\
 PO Box 210011, Cincinnati, OH 45221, U.~S.~A.\\  

  E-mail: \email{hrb@slac.stanford.edu}}

\abstract{
The Daya Bay reactor antineutrino experiment is designed to make a precision measurement of the neutrino mixing angle $\theta_{13}$, and recently made the definitive discovery of its non-zero value. It utilizes a set of eight, functionally identical antineutrino detectors to measure the reactor flux and spectrum 
at baselines of $\sim$ 300 - 2000m from the Daya Bay and Ling Ao Nuclear Power Plants. The Daya Bay antineutrino detectors were built in an above-ground facility and deployed side-by-side at three underground experimental sites near and far from the nuclear reactors. This configuration allows the experiment to make a precision measurement of reactor antineutrino disappearance over km-long baselines and reduces relative systematic uncertainties between detectors and nuclear reactors. This paper describes the assembly and installation of the Daya Bay antineutrino detectors.
}


\begin{document}

\section{Introduction}

Neutrino mixing  was first established by the Super-Kamiokande,~\cite{KAMIOKANDE}, SNO~\cite{SNO}, and KAMLAND~\cite{KAMLAND}, experiments using neutrinos from the atmosphere, the Sun, and commercial nuclear reactors. The observation of neutrino mixing implies a non-zero neutrino mass and the difference in the mass eigenstates yields a characteristic oscillation length for neutrino oscillation. In analogy with quark mixing, neutrino mixing is described by the Pontecorvo-Maki-Nakagawa-Sakata (PMNS)  matrix~\cite{PMNS1,PMNS2} relating the three neutrino flavor and mass eigenstates.  The PMNS matrix is parameterized by three mixing angles and  complex phases. Two of the three mixing angles had been measured in the solar and atmospheric neutrino experiments, but the neutrino mixing angle $\theta_{13}$ was unknown until the recent   reactor antineutrino measurements by Daya Bay \cite{tdr,DB}, Double Chooz \cite{DC}, and RENO \cite{RENO}. In 2012, the Daya Bay Reactor Neutrino Experiment discovered non-zero $\theta_{13}$ and made the most precise measurement of this mixing angle to date \cite{DB,DB2}. The measurement of $\theta_{13}$ establishes neutrino oscillations over km-long baselines for reactor neutrinos with energies up to $\sim$8~MeV.

The precision measurement of reactor antineutrino disappearance in Daya Bay (and RENO) is made possible by comparing  the reactor antineutrino flux 
at several different distances from the reactor cores and making a relative measurement between detectors near and far from the nuclear reactors.  Daya Bay's most recent measurement of
 $\sin^2 2\theta_{13}$ $=0.089\pm 0.010~\rm{(stat}) ~\pm0.005 (\rm{syst})$
is still dominated  by the statistical error in the number of antineutrino events observed at the far site \cite{DB2} since the carefully designed AD assembly has kept systematic errors very low.
However, with several more years of data this statistical error  will decrease and become smaller than the 
present systematic error.  The goal is to minimize all possible systematic errors to reach the ultimate sensitivity of the experiment. 

A relative measurement between near and far detectors allows for the cancellation of many common systematics. Understanding the relative detection efficiency though is key to the success of the experiment. The Daya Bay experiment has used an extensive suite of techniques to make its eight detectors as identical as possible. This includes amongst others the construction and assembly of all eight  detectors
in the same clean room with an identical set of  procedures and  filling the detectors with liquids
from the same storage tanks to ensure identical behavior of the detector liquids. 

The acrylic and stainless steel structures of the Daya Bay antineutrino detectors are not built to sub-mm tolerances by normal construction techniques and small variations between detectors from differences in the fabrication and assembly are to be expected.
To track possible differences between detectors  detailed surveys of the as-built detectors  were made to characterize 
each of them during the construction and assembly process. 
Simulations and detailed studies with in-situ data established that the detectors are functionally identical.
Moreover, the Daya Bay detectors are interchangeable and can be moved from one experimental hall to another if desired.

A side-by-side deployment and comparison of the first two Daya Bay detectors~\cite{DB3}  has shown that  detected antineutrino rates 
were  identical to within 0.2\%. This paper describes in detail the assembly and 
installation of the Daya Bay antineutrino detectors and the steps taken to control the detector-to-detector variation.
       
\subsection{Overview of Daya Bay Reactor Neutrino Experiment}

The mixing angle $\theta_{13}$ is a fundamental parameter of nature. Precise knowledge of $\theta_{13}$ is
needed to plan future experiments designed to measure the neutrino mass hierarchy and CP violation in the neutrino sector.
Daya Bay uses electron-type antineutrinos from six high power (2.9 GW$_{th}$)
commercial nuclear reactors to measure a deviation from the expected $1/r^2$ behavior in the
number of antineutrino interactions observed as a function of distance from the nuclear reactor
cores to the detectors. 

Antineutrinos are detected by interactions in the hydrogen-rich scintillating liquid target  via the inverse beta-decay reaction (IBD), 
\begin{equation}\label{eqn:IBD}
\overline{\nu}{_e} + p \rightarrow e^+~+~n.
\end{equation}

The positron annihilates in the liquid scintillator producing a prompt energy pulse. 
The neutron thermalizes and is then absorbed by a gadolinium(hydrogen) nucleus with a average neutron capture lifetime of
$\sim~30(200) \mu$sec producing an
delayed energy pulse. This characteristic time-correlated energy signal allows the antineutrino
signals to be cleanly separated from the much  higher rate ($\approx10^4$) radioactive backgrounds.

\begin{figure}\hfil
\includegraphics[clip=true, trim= 15mm  75mm 15mm 15mm,width=6.0in]{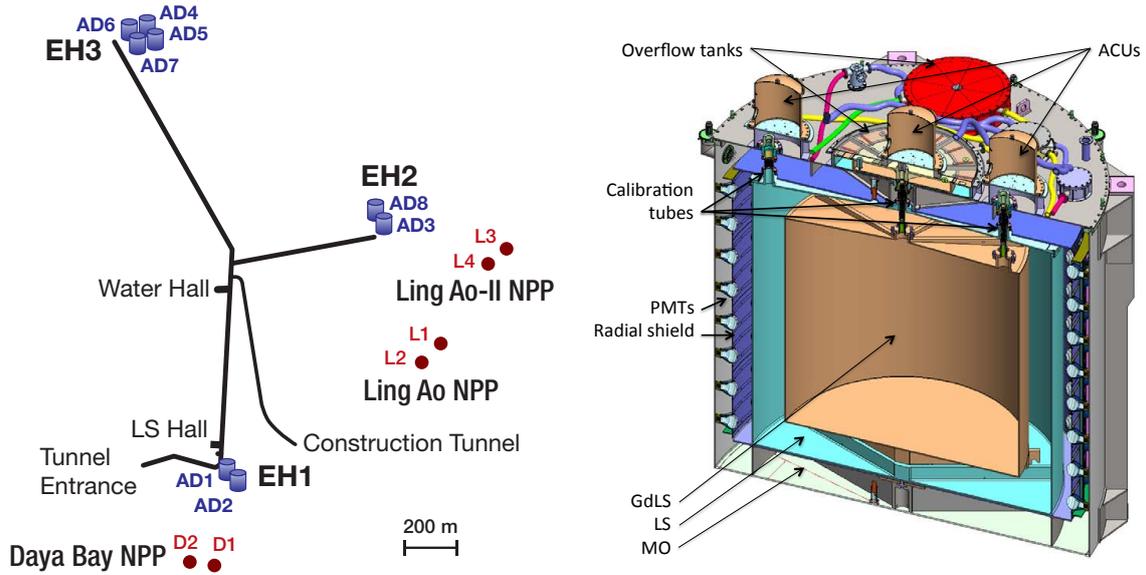}\hfil
\caption{(left) Layout of the Daya Bay experiment. The dots represent reactors cores, labeled as D1, 
D2, L1, L2, L3 and L4. There is one pair of cores at each of the Nuclear Power Plants (NPP). 
Eight antineutrino detectors (ADs) are installed in three experimental halls.
(right) The Daya Bay antineutrino detector contains 3 zones: a gadolinium loaded liquid 
scintillator inner target
(20 tons) inside the inner acrylic vessel (brown), a liquid scintillator gamma catcher (21 tons) 
contained by the outer acrylic vessel (aqua), and a mineral oil buffer (37 tons) inside the stainless steel outer vessel (grey).
}\label{fig:Int_8AD}
\end{figure}

As shown in Fig.~\ref{fig:Int_8AD} the Daya Bay experiment consists of three underground experimental halls 
located 360-2000 m  from three pairs of reactor cores. The experimental halls are excavated 
under the side of an adjacent mountain to reduce cosmic ray backgrounds and are connected by large aperture tunnels.
Each of the two near experimental halls contains two antineutrino detectors (ADs) in an instrumented water pool. 
The far experimental hall contains four ADs in a  larger water pool. The ADs are constructed in an above ground assembly building
before being transported underground to the liquid scintillator (LS) filling hall.  The filled ADs are then moved on a custom
transport to the appropriate experimental hall and lifted by an overhead crane into the water pool.

Each AD consists of a Stainless Steel Vessel (SSV) containing 3 detector zones filled with
different liquids as shown in Fig.~\ref{fig:Int_8AD}. Two nested acrylic cylinders separate the 3 zones. The
innermost target zone contains 20 tons of gadolinium-doped liquid scintillator (GdLS) inside an
Inner Acrylic Vessel (IAV). This zone is surrounded by 21 tons of liquid scintillator (LS)
gamma catcher contained by an Outer Acrylic Vessel (OAV). The scintillation light from IBD interactions
is observed by 192 photomultipliers (PMTs) arranged in a cylindrical shell. An outer zone of
liquid contains 37 tons of mineral oil (MO) that shield the inner zones from radioactivity originating in the 
PMT glass or other background sources. Overflow tanks above the SSV lid allow for the 
expansion of the liquids with changes of temperature or external pressure.
Reflectors above and below the LS volume iimprove the light collection efficiency, 
increase the uniformity of the light collection, and reduce variations in 
the energy response of the detector along its vertical axis.
Three automated 
calibration units (ACU) above the SSV lid
allow remote deployment of radioactive sources or LEDs into the GdLS or LS liquid
volumes through three calibration tubes. Weekly calibration data runs are interspersed with normal data taking to accurately measure
and track the energy response of each AD.

The antineutrino detectors sit in water pools which shield the detectors  in all directions
from ambient radioactivity with > 2.5 m of water.
The water pools are instrumented with PMTs arranged in optically
separated inner (IWS) and outer water (OWS) pool zones to detect muons which may 
introduce  spallation neutrons or other cosmogenic backgrounds into the ADs. 
Additional muon detection is provided by four layers of resistive plate chambers (RPCs) which are rolled over the water pool.

\begin{figure}\hfil
\includegraphics[clip=true, trim=10mm 60mm 10mm 25mm,width=6.0in]{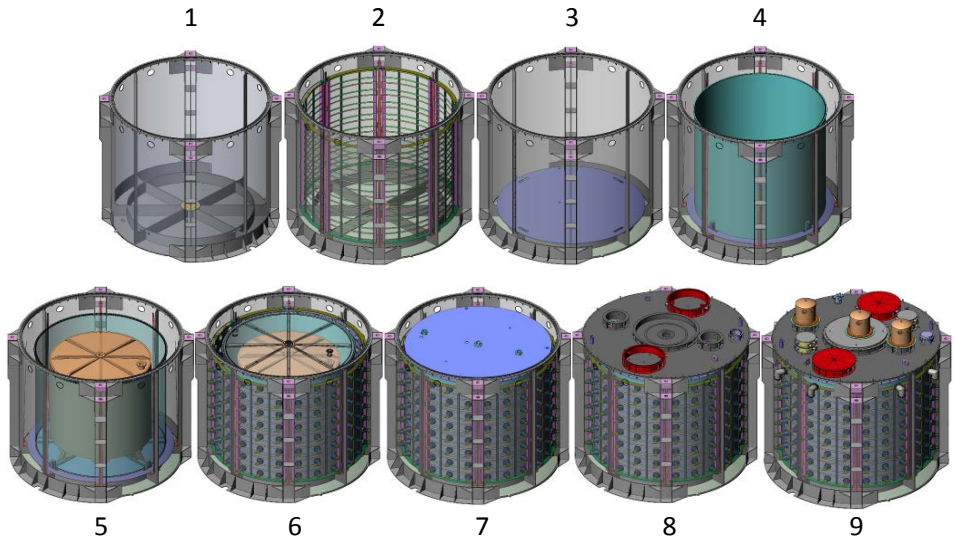}\hfil%
\caption{AD assembly sequence. (1) After cleaning, the SSV is placed in the assembly pit. 
(2) Bare PMT ladders are test mounted and surveyed. (3)The bottom reflector is inserted. 
(4) The OAV is placed on the reflector. (5) The IAV is brought in and mounted inside the OAV. 
(6) the OAV lid is mounted and the PMT ladders are installed. (7) The upper reflector and lower parts of the OAV calibration tubes are installed. 
(8) The SSV lid is installed. (9) The  overflow tanks, upper calibration tubes, covers and ACUs are installed. }\label{fig:Const_seq}
\end{figure}

ADs were constructed and assembled above ground in parallel with tunnel and experimental hall excavation to 
start the experimental program as quickly as possible. 
The nested design of the antineutrino detector required a complex assembly sequence as shown in 
 Fig.~\ref{fig:Const_seq}.  Components were surveyed and leak checked at each step before access was blocked by
 the installation of the next component.  All of the AD instrumentation was tested without liquids  upon completion of the assembly 
to ensure that the AD was ready to be filled.
Completed "dry" detectors weighing $\approx~30$ tons were transported underground 
through a 10\% grade entry tunnel and then filled with detecting liquids. The filled detectors now weighing  $\approx~110$ tons
were then moved along the flat  underground tunnels (0.5\% grade) to the experimental halls  and were lifted into place by large overhead cranes.

\subsection{Antineutrino Detector Design}

The detailed antineutrino detector design is driven by the requirements that the ADs must function reliably over five years,
be moveable, 
have three entirely filled liquid zones,  have multiple calibration positions, 
and be immersed in meters of water.
To fully exploit the reduction of systematic errors possible with measurements at multiple baselines
the relative acceptance and efficiency of the ADs should be known  with a 
precision similar to the proposed statistical error at the far site (0.2\% in 3 years). 
A high statistics comparison of the detected antineutrino rates at each baseline  will 
be used to verify the AD acceptance and efficiency calculations.

\begin{figure}\hfil
\includegraphics[clip=true, trim=0mm 0mm 0mm 0mm,width=5.5in]{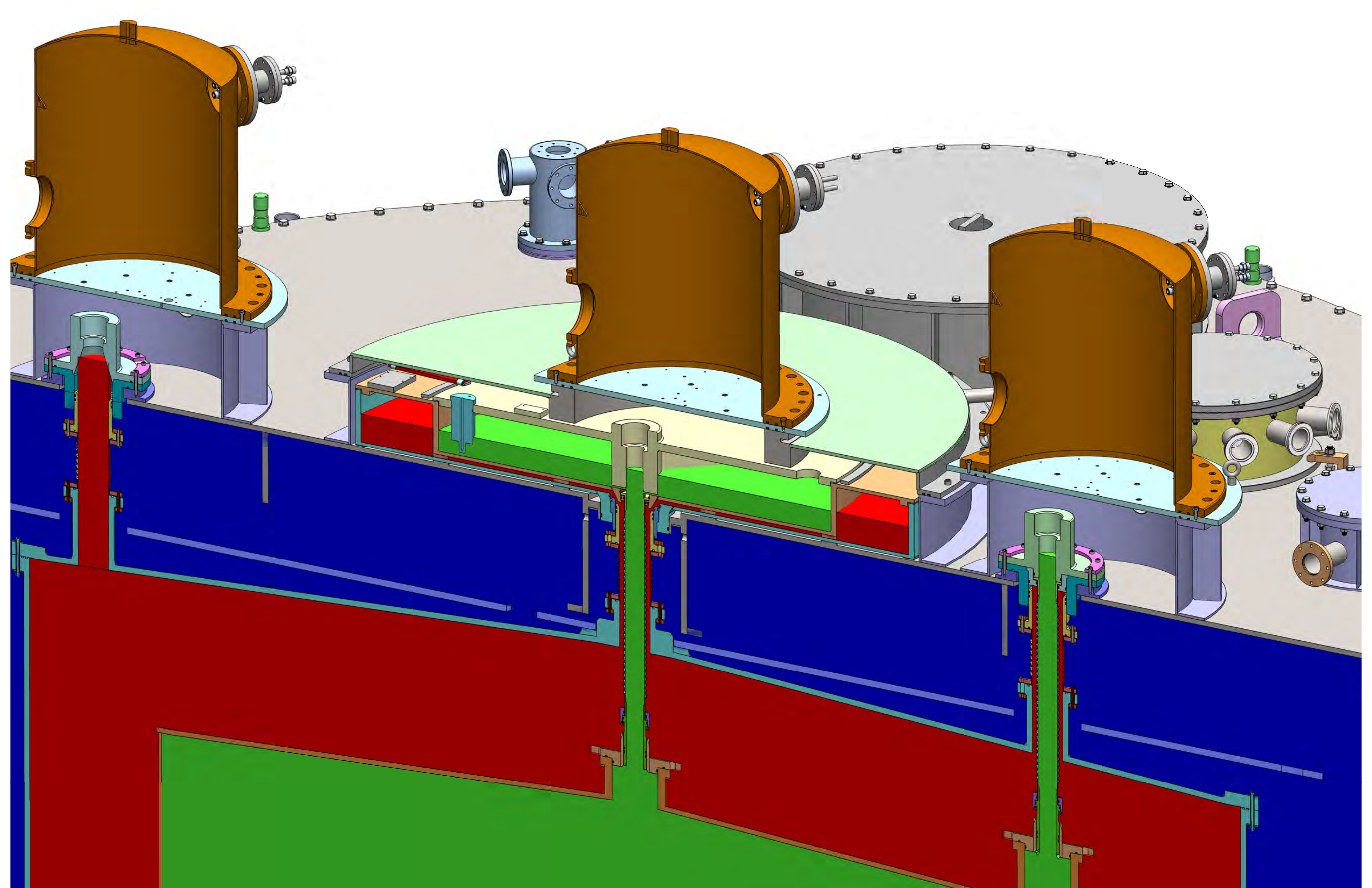}\hfil%
\caption{Liquid volumes inside  the Daya Bay  vessels, calibration tubes, and overflow tanks: GdLS - green, LS-red, MO-blue.
The outside of one of the MO overflow tanks is seen  behind the central ACU. The IAV and OAV are connected by convoluted teflon tubes to the SSV lid or to the central
overflow tanks. All surfaces contacting the GdLS or LS (red or green) are constructed from acrylic or teflon components. }
\label{fig:Liquid}
\end{figure}

AD liquids need expansion volumes to allow for temperature changes during transport, storage, 
and water pool temperature variations. Liquid levels also change as the entire SSV is compressed during the 
filling of the water pool.
Overflow tanks were positioned on the SSV lid as shown in Fig.~\ref{fig:Liquid}. This also keeps
the internal volumes full so there is no sloshing during transportation and rigging.  Teflon and acrylic calibration tubes 
which connect the overflow tanks and automated calibration units to the acrylic vessels containing the detector liquids
are also shown.   The flexibility of the 
convoluted teflon tubes and sliding O-ring seals allow for small misalignments or movements of the acrylic vessels
during construction or transport. 

\begin{figure}\hfil
\includegraphics[clip=true, trim=0mm 10mm 0mm 20mm,width=5.5in]{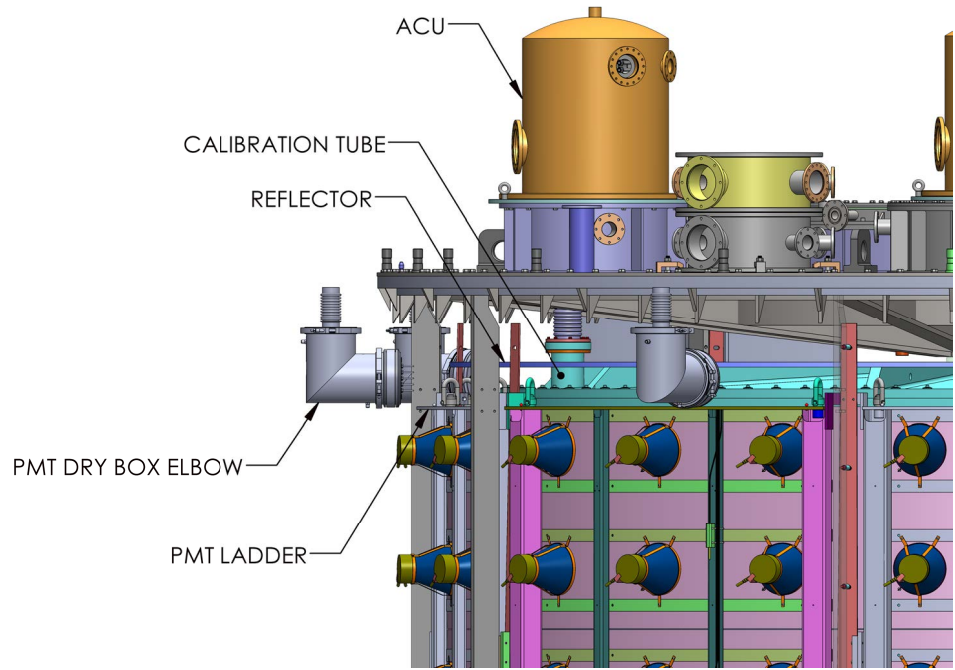}\hfil%
\caption{AD section with the SSV wall removed to show details of the PMT mounting method, the dry box connections,
upper reflector support and one of the OAV calibration tubes. Above the SSV lid are ACUs, electrical interface boxes
and overflow tanks. The complicated geometry of the SSV ribs and tube connections are hidden from the PMT field of view by the reflector. }
\label{fig:AD2}
\end{figure}

Daya Bay ADs have reflectors below and above the
GdLS target volume to reduce the number of PMTs required and to simplify construction (see Fig.~\ref{fig:AD2}). 
Monte Carlo simulations~\cite{tdr} predict an acceptable light collection uniformity. The reflector also has the advantage of
hiding complex lid or SSV base geometry from the PMT's field of view.

The vessels containing the inner detector liquids are made from cylinders of UV transparent acrylic~\cite{AV,AV2}. 
Cylinders are easier and cheaper to manufacture than spheres, stronger than cubes and work well with end reflectors. Acrylic has a similar refractive index 
as GdLS and LS and is chemically compatible.  When filled with liquids the acrylic vessels are optically transparent.
Acrylic is also chemically pure, containing low natural radioactivity, can be made into thin walled structures, and has been successfully used in previous neutrino detectors.  However, long term stresses in acrylic can cause crazing and cracking 
of the material and must be managed carefully.

To speed construction and to protect the PMTs from potential damage when the large and bulky acrylic vessels were installed 
into the SSV, PMTs were pre-mounted on sub-assembly frames holding 24 PMTs and an opaque radial 
shield (PMT ladder) which could then be installed after the OAV and IAV.  
This allowed the PMT ladders to be prepared in parallel with the OAV and IAV installation, survey and leak checking.
The PMT cables were routed to liquid-tight feedthroughs that were mounted on the SSV wall. Both the feedthroughs and electrical 
connections could be prepared and tested before installation into the AD.

\section{AD Assembly}

The large components of the ADs were constructed at factories in Taiwan~\cite{Nakano}, China and the U.S.~\cite{Reynolds}
and shipped to Daya Bay for assembly. 
These components were then cleaned before being brought into the assembly clean room.
Two ADs were assembled in parallel over a period of 4-12 months and filled sequentially within two weeks.
This pairwise assembly sought to reduce differences between assembled and filled AD pairs by exposing the ADs to the 
same construction environment and filling with similarly aged scintillating liquids. The first
AD pair in Experimental Hall 1 were found to be functionally identical~\cite{DB3} .  
All six of the first ADs have performed similarly, relieving the need to deploy ADs 
with one  of each pair in a near and far experimental hall as originally planned. This option is still available if 
warranted by remaining systematic uncertainties.

Full-sized prototypes of the outer (OAV) and inner (IAV)  acrylic vessels were constructed  and shipped to
Daya Bay.  These  and other prototype parts were used to fully debug the assembly sequence and procedures.
All critical lifts were practiced and leak checking procedures developed. The prototype assembly  was an invaluable learning 
period  and speed assembly of the eight detectors.

\subsection{Assembly Facilities and Fixtures} 
 
\begin{figure}\hfil
\includegraphics[clip=true, trim=35mm 86mm 18mm 55mm,width=6.0in]{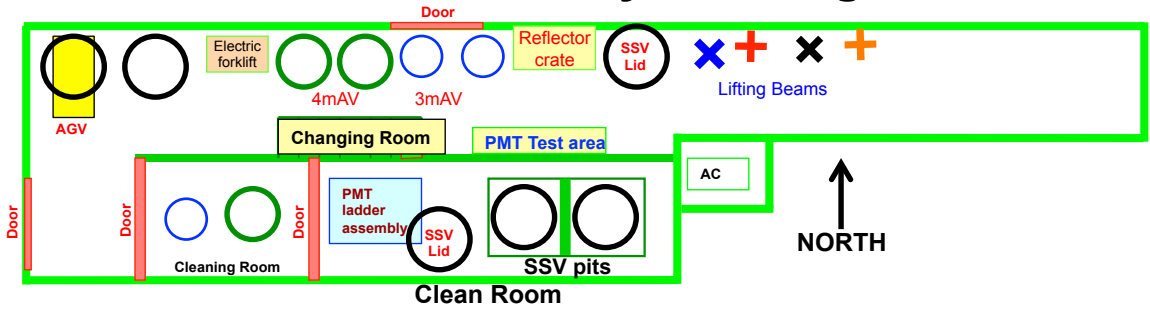}\hfil%
\caption{Surface Assembly Building during the height of activity. Two partially assembled stainless steel vessels (SSVs -black circles) are in the clean room.
A recently cleaned 4 meter outer acrylic vessel (OAV - green circle) waits  in the cleaning area.  A  3 meter inner acrylic vessel 
(IAV - blue circle) is also being prepared for cleaning.  Other components are stored in the neighboring bay.}
\label{fig:SAB}
\end{figure}

The ADs were assembled in a dedicated above ground building inside a large clean room
 before transportation underground for filling and installation. A plan view of the Surface Assembly Building (SAB) is shown in Fig.~\ref{fig:SAB}.
The north side of the building  was serviced by a 10-ton overhead crane and used primarily as a storage area.
The south side of the building was underneath a 40-ton overhead crane and contained a 11~m $\times$ 33~m $\times$ 11 m 
class-10,000 clean room and a smaller adjacent cleaning area. 
Particle counts in the clean room were measured weekly and were typically $\leq 1000$ counts/cc.
The clean room had two large pits 4 m deep into which SSVs were placed at a convenient working height.  
The crane clearance was chosen to allow the transport of a SSV  into the far pit even if the close pit was occupied
thus providing more than sufficient clearance to install acrylic vessels
 or PMT ladders. A section of the clean room was reserved for the assembly and storage of the PMT ladders.  
Other areas were used to store
SSV or OAV lids that had been removed from the AD for access to the interior. 
A gas rack was located between the SSV pits to supply high flow rates of argon or freon at low pressure for leakage tests.

The clean room was connected by a roll-up curtain wall  to the cleaning area.  
This semi-clean area was isolated from the rest of the SAB by another roll-up curtain wall which kept
particulate levels low.
The cleaning area was equipped with a resin bed water filtration system which provided > 10 megaohm-cm  water for cleaning
items that would enter the clean room. 

Although the cleaning area was nominally under the 40-ton crane coverage, use of the crane required 
that the roll-up curtain be raised to the ceiling between the rooms so that the crane could be moved into position,
reducing the clean room air quality.  In addition there were  numerous occasions  during the prototype AD assembly when
a crane was needed in both areas. To address these concerns a second 10-ton crane was mounted in the washing area.
Large items such as a SSV or AV were transported between the cleaning area and clean room on an Automated Guided Vehicle (AGV).  

\begin{figure}\hfil
\includegraphics[clip=true, trim=0mm 20mm 0mm 0mm,width=6.0in]{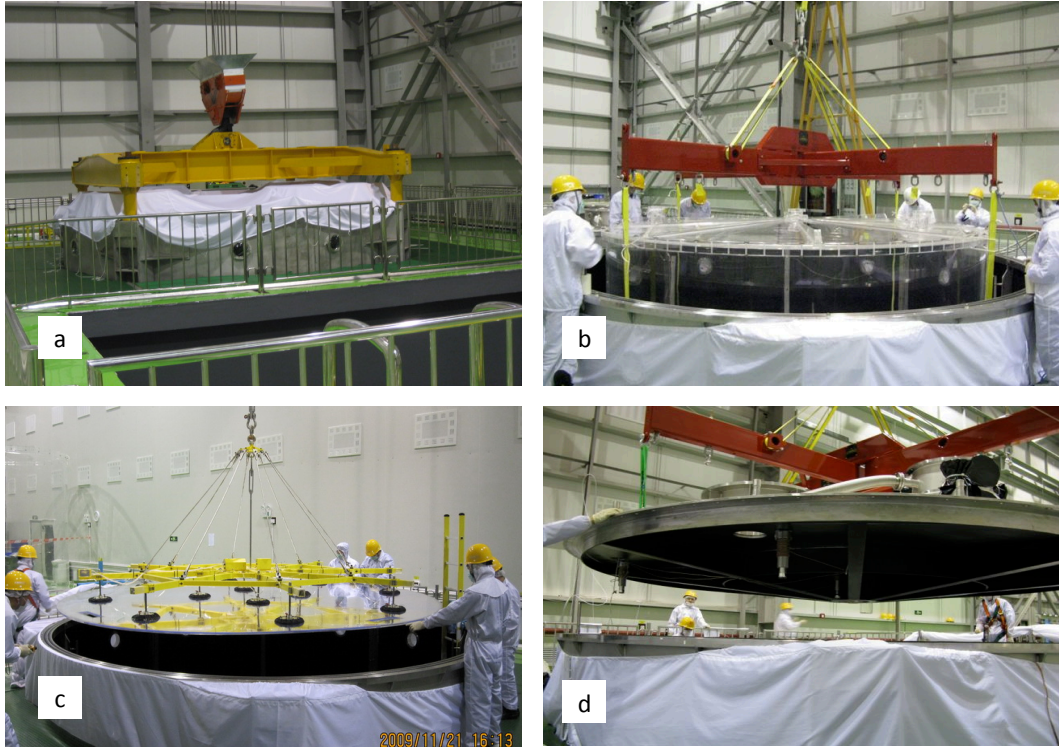}\hfil%
\caption{Custom lifting fixtures for AD assembly. (a) SSV lifting fixture (yellow) connected to a SSV.
(b) multipurpose fixture(red) used to lower an OAV into the SSV.
(c)  vacuum pad  fixture (yellow) used to install the reflector. (d) the multipurpose fixture 
lowering the SSV lid into place.}
\label{fig:LIFT}
\end{figure}

Several custom lifting fixtures were designed and built to move AD components safely. 
Two  fixtures were designed to lift a SSV.
Both fixtures were of a low profile design to minimize height requirements in the SAB and experimental halls.
Each fixture has  four lift points at a 5400mm diameter outside the SSV cylinder.
At each lifting point, a M100 screw rod on the lifting fixture is inserted into a hole on SSV and engaged into a floating nut.
The smaller  lifting fixture, shown in Fig.~\ref{fig:LIFT}a,  is rated for 32 tons and is used to move SSVs or ADs in the SAB.
The larger 110 ton fixture, shown in Fig.~\ref{fig:InstallB},  is reserved for lifting filled ADs in the underground halls.

A  lighter more general multipurpose fixture,  shown in Fig.~\ref{fig:LIFT}b and  Fig.~\ref{fig:LIFT}d, 
was used to lift lighter components such as the acrylic vessels or the SSV lid.  
This X-shaped fixture had lifting eyes at  diameters corresponding to the OAV and IAV and was connected by slings to the bottoms of the acrylic vessels.
Lifting the acrylic vessels from the base minimized stresses (and thus required thickness) in the cylindrical walls but
required a custom sling connection  to the AV lifting ears which could be disconnected when the lifting ears were at the bottom of the SSV.
The connection was made from a rectangular frame. 
The bottom of the frame hooked beneath the lifting ear.
The top of the frame was connected to the lifting sling.
A third cross piece near the center of the frame supported a  pusher block which locked the lifting frame in place when tension
was applied to the sling.
After the lift when  the sling was slack, a  rope was used to lift the block and with further travel upward pushed the lifting frame clear of the lifting ear.  

Lifting the thin, large flat reflectors into the AD required another custom lifting fixture shown in Fig.~\ref{fig:LIFT}c.
Twelve suction cups were supported on a steel spider and connected to a common vacuum pump.  
When evacuated the fixture could pick up the 400 kg reflector deliver it to the bottom of the SSV and then by releasing 
the vacuum, disengage from the reflector. 

Multiple platforms were needed in the assembly room to provide  access to the interior of the 5~m wide and 5~m deep AD.
Semi-permanent steel platforms fitted snugly around the SSV covering the pits. 
Two large half-circular platforms could be put onto the SSV flange
for installation of the calibration tubes to the OAV ports. 
Additional attachments  gave access to the lower IAV ports for leakage tests as seen in Fig.~\ref{fig:Leak}.
Smaller platforms could be lowered onto the OAV ribs to support workers during placement and survey of the
IAV shims.   

\begin{figure}\hfil
\includegraphics[clip=true, trim=25mm  30mm 25mm 50mm,width=5.0in]{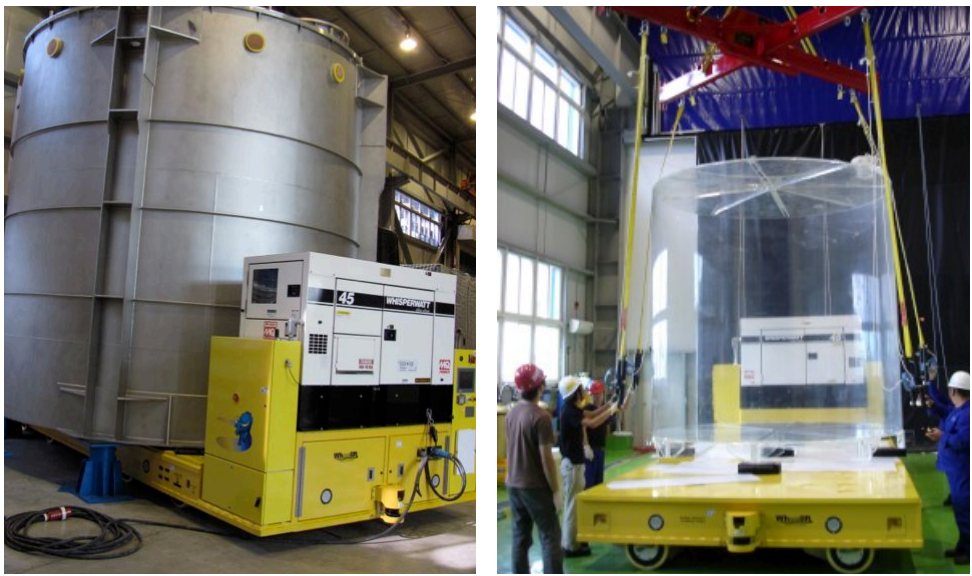}\hfil%
\caption{(left) The SSV is being lifted by the AGV from  storage stands prior to transport to the cleaning room. (right) 
An IAV has been lifted onto the AGV after cleaning.
 }\label{fig:ASS_AGV}
\end{figure}

An Automated Guided Vehicle (AGV) was specifically designed to transport completed ADs down the 10$^\circ$ slope 
to the LS hall and filled ADs weighing 110 tons to the experimental halls. 
The AGV proved to be a versatile workhorse in moving SSVs and acrylic vessels into and 
out of the assembly areas as shown in  Fig.~\ref{fig:ASS_AGV} since it could operate under diesel or electric power.

\subsection{Assembly Sequence} 

Assembly of an AD followed a well defined sequence of steps,  previously shown in Fig.~\ref{fig:Const_seq} 
and described below.

\begin{enumerate} \itemsep1pt
\item After the  SSV was cleaned in the wash area it was  brought into the clean room and lowered into the pit onto
leveled stands. 
Platforms were installed around the SSV.  
A survey of points on the SSV lid established a coordinate reference 
frame centered on the central port in the lid.
The SSV lid was then removed for access to the interior.
\item 
A standard set of points along the bottom ribs and cylinder inside the SSV were surveyed. 
PMT ladder rails were installed and clean PMT ladders  were test mounted in the SSV. 
The ladder positions were surveyed and checks were made on the fit of the PMTs.
After survey the ladders were removed and populated with PMTs.
The PMT ladder assembly will be covered in Section 2.4.
\item
Three 5-cm monitoring PMTs were mounted on the SSV floor and surveyed.
A copper disc used to shield radiation from the SSV center rib welds  was installed. 
The reflector was  lowered to rest on the SSV ribs and locations on its surface were surveyed.
\item
A clean outer acrylic vessel (OAV) with attached lid was brought into the clean room and lowered  onto the reflector. 
The OAV was surveyed and the center port  aligned to the SSV center. 
The OAV was  then rotated to best align the outer ports and  bolted to the SSV.
The OAV lid was then removed. 
\item
The inner acrylic vessel (IAV) support pucks were mounted on the OAV ribs and shimmed level. 
After the IAV was cleaned, it was installed into the OAV. 
The position of the calibration ports were surveyed.
If IAV ports were not aligned with the SSV openings  the IAV was removed and additional shims
were added to move the top of the IAV as required.
The IAV hold-downs were lifted into position, securing the IAV to the OAV.
Access platforms were installed over the center of the AD to install the 
IAV port covers.
The IAV calibration tubes  were   temporarily  installed into the port covers and connected to the 
gas rack with large diameter gas lines.
The IAV was filled with > 75\% argon at 12 cm of water equivalent pressure. An argon sniffer was used to check
the IAV port O-ring seals.
\item 
Double O-rings were put into the grooves on the OAV flange before the OAV lid was reinstalled.
Ninety six bolts were used to evenly clamp the lid to the OAV flange. 
A pressure test verified the status of the double O-ring seal. 
Eight completed PMT ladders were installed into the SSV and  the PMT feedthrough flanges were mounted on the SSV wall. 
Each feedthrough flange/cable assembly was leak checked.  
Bad feed-throughs or bad cables were replaced by removing the ladder
and swapping the PMT/cable assembly. 
After all leak checks were completed,
PMT functionality tests were performed with a temporary cover over the AD.
Any PMT failing  tests was replaced by removing the ladder and swapping out the failed PMT.
Similarly, the AD cameras, temperature sensors, and LEDs were tested and replaced if necessary. 
\item
The upper reflector was mounted on the OAV lid.
The OAV calibration tubes were installed on the OAV port. Test fittings connected the OAV tubes to the gas rack
and were used to leak test the OAV calibration tube connections. 
\item
The SSV lid was installed and all
bolts torqued to specification. 
The double O-rings of the SSV lid were leak checked.
The off axis calibration tubes were connected to the SSV lid. 
The LS overflow tank was installed  and the central calibration  tube connected.
A leak test of the three liquid volumes MO, LS, and GdLS was performed.
\item
The LS overflow tank, lid, and
lid sensors ~\cite{Mass} were installed in the central overflow region.
Cover gas distribution lines were installed  above the overflow tanks  and other gas volumes on the SSV lid and connected to a central gas 
manifold on the SSV lid~\cite{Gas} .
The central overflow and MO overflow lids were  installed.
Cameras monitoring the liquid level in the off-axis calibration ports were installed.
The MO clarity system was installed and cabled.
Cables from the LED calibration system were run through feedthroughs.
ACUs were mounted and connected to temporary control electronics for electrical and mechanical deployment tests. 
\end{enumerate}

\subsection{Cleaning} 

Both the interior and exterior of each AD were thoroughly  cleaned to reduce possible radioactive contamination 
of the numerous surfaces and to avoid residues which could interact with the AD liquids. 
Since a few grams of sweat or dirt contain $\sim$10\% of the expected activity of twenty tons of GdLS  target liquid, 
AD assembly was performed inside a clean room with all workers wearing clean room suits.
Since all of the larger AD components were trucked and/or shipped by sea to Daya Bay with varying levels of 
protection from the outside elements, standardized cleaning procedures were applied to all components 
entering the clean room.

Large items were unpacked in the storage areas of the SAB before being brought to the cleaning area.
Smaller items were cleaned in ultrasonic cleaners in two steps, first  with a 1\% Alconox solution and then with
a 10 megaohm-cm water bath.

\begin{figure}\hfil
\includegraphics[clip=true, trim=10mm  140mm 10mm 10mm,width=6.0in]{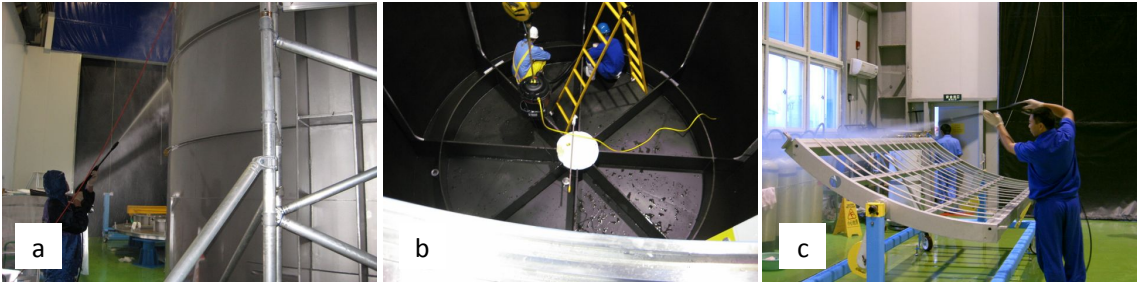}\hfil%
\caption{ (a) Pressure cleaning of the outside of the SSV. (b) Cleaning the interior of the SSV.
(c) Pressure cleaning of a PMT ladder..}\label{fig:Clean1}
\end{figure}

A summary of the SSV cleaning procedure follows.
The  SSV was delivered into the cleaning area and set on temporary stands.
The SSV lid was removed and stored for a separate cleaning.
The top flange and grooves were cleaned with a microfiber cloth and a 1\% Alconox solution.
The flange area was then pressure rinsed with high resistivity water.
Cleaning the outside of the SSV followed the same procedure as seen in Fig.~\ref{fig:Clean1}a.   
Extra spot scrubbing was done with $Al_2O_3$ powder  and  microfiber cloth. 
The rinse water was collected and measured 
with a conductivity meter. When the rinse water conductivity matched that of
the water  from the pressure washer, $\approx ~6\mu S$, rinsing was considered complete. 
The cleaning of the interior of the SSV, shown in Fig.~\ref{fig:Clean1}b, followed similar steps. A drain plug was opened at the
bottom of the SSV to let most of the rinse water escape. Wet/dry vacuums help to clear any
residual puddles. The SSV lid was washed in a similar manner before being repositioned on the SSV.

\begin{figure}\hfil
\includegraphics[clip=true, trim=10mm  140mm 10mm 10mm,width=6.0in]{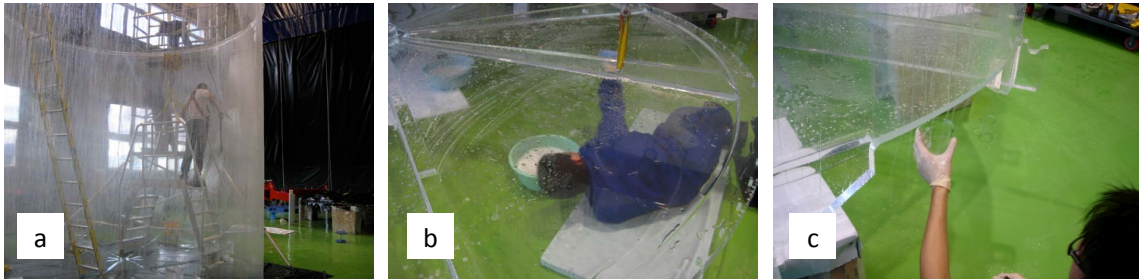}\hfil%
\caption{ (a) Pressure cleaning of the inner OAV. (b) Scrubbing the bottom of an OAV with an Alconox solution.
(c) Collecting the rinse water from an IAV. 
}\label{fig:Clean}
\end{figure}

The large acrylic vessels were cleaned in a similar manner. Only the exterior of the IAV was cleaned 
in the SAB since the interior had been  cleaned at the factory~\cite{Nakano} before the final bond
of the top and bottom acrylic pieces.  The lid of the OAV was removed so that both the interior and exterior
could be cleaned. Photographs of OAV and IAV cleaning appear in Fig.~\ref{fig:Clean}.
Some OAVs  had a residue from the protective film applied before shipping that was particularly troublesome
to remove. In the worst cases the residue was removed with a copper scrubber which left the surface lightly
scratched. This treatment was not expected to alter light transmission through the acrylic once the the OAV was immersed
in MO. More details of the acrylic vessel cleaning can be found in \cite{AV}.

The PMT ladders displayed some corrosion after shipment to Daya Bay from the United States.
An additional acid cleaning step at a local factory was added to the cleaning regimen
with an extra vigorous pressure wash as seen in Fig.~\ref{fig:Clean1}c. 

\subsection{Assembly of the Photomultiplier Ladders} 

After the PMT ladders were test fitted and surveyed, the ladders were removed from the SSV
and stored in the clean room.  
In parallel with the installation of the acrylic vessels the ladders were populated
with 24 PMTs and other sensors. 
 PMTs (Hamamatsu R5912~\cite{Hamamatsu} with low radioactivity glass) had been previously cleaned and tested at an off-site facility and
required only light cleaning with a microfiber cloth.  
Other components of the PMT ladder assembly had been previously cleaned
by the usual procedures described in Section 2.3.

\begin{figure}\hfil
\includegraphics[clip=true, trim=20mm  10mm 20mm 0mm,width=5.0in]{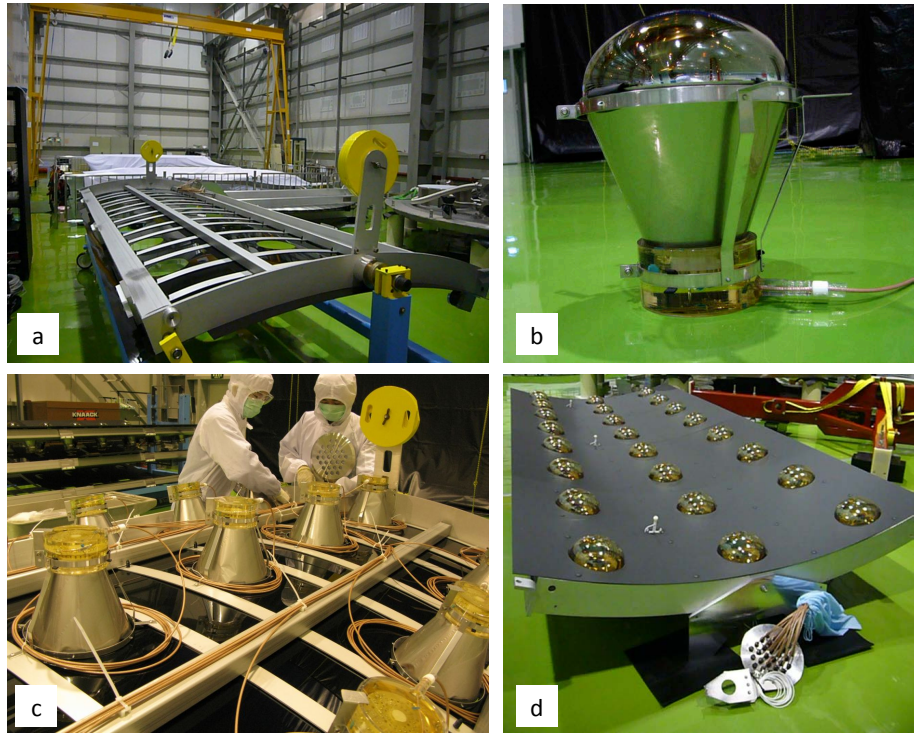}\hfil%
\caption{ (a) Empty ladder mounted on assembly frame.  (b) PMT with magnetic shield ready to mount.
(c) Routing signal cables after PMTs were mounted. (d) Completed PMT ladder with calibration LEDs installed.}\label{fig:Const_PMT}
\end{figure}

The PMT ladder was first mounted on an assembly frame as shown in  Fig.~\ref{fig:Const_PMT}a.
A 3.2 mm thick matte black acrylic radial shield was attached to the ladder frame.
Some PMT ladders had extra components. Two ladders in each AD had a camera~\cite{Camera} to view
the interior of the AD and to assist in AD filling. 
Other ladders had  three calibration LEDs,  or three MO temperature sensors, 
or a sealed retroreflector for the MO clarity system. These extra items were mounted
before the more fragile PMTs were installed.

Next the PMTs were prepared  by wrapping them in a FINEMET magnetic shield~\cite{FINEMET} and placing them in the 
stainless steel PMT mount.
Short Viton U-channel pads were used to cushion the mount  where it touched the PMT glass or base.
A completed PMT assembly is shown in Fig.~\ref{fig:Const_PMT}b.
Each 7-m-long RG303 cable carrying both signal and HV was tagged with the PMT serial number and location on the ladder.
The PMT assemblies were then bolted to pre-drilled holes in the ladder frame. 

Cables were routed along and secured to the ladder frame (Fig.~\ref{fig:Const_PMT}c) with the cables sticking past the
top of the frame by 1 m. The ends of the PMT cables had been previously terminated in a feedthrough plug and SHV connector
by Hamamatsu.
The PMT cable was threaded through the specified  hole in the feedthrough flange and snugly fastened by a retaining ring.
As the number of cables varied with the number of extra components mounted on the ladder any unused holes were sealed
with blank plugs. A completed PMT ladder with cables mounted on the feedthrough flange is shown in Fig.~\ref{fig:Const_PMT}d.

\begin{figure}\hfil
\includegraphics[clip=true, trim=0mm  45mm 0mm 50mm,width=5.0in]{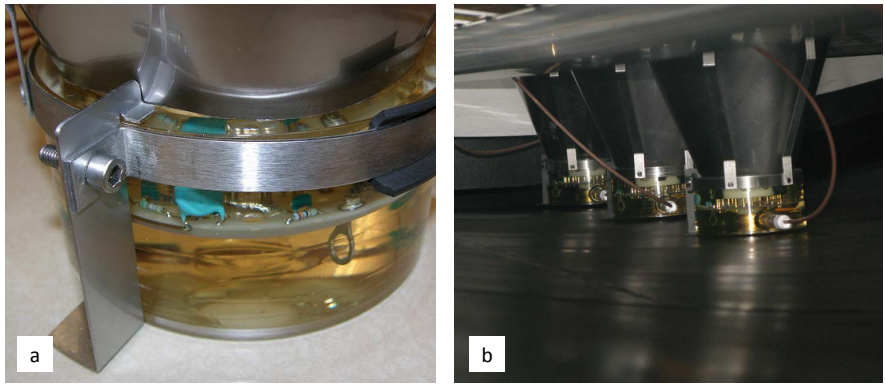}\hfil%
\caption{ (a) Details of the PMT mount showing the retaining clamp which can be adjusted to move the PMT radially in or out
as needed.  (b) View of an installed PMT ladder showing  typical clearances.
}\label{fig:PMT2}
\end{figure}

One minor modification of the original design was required. Both the PMT frames and SSV cylinder deviated from true cylinders
by up to 35 mm. In some cases the back end of the PMTs touched the SSV inner wall. A retaining clamp was added to the design
which could be adjusted to move the PMTs radially inward. PMTs were moved inward up to 20 mm  from their design position when necessary. 
A closeup of the clamp and PMT base is seen in Fig.~\ref{fig:PMT2}a.  
Fig.~\ref{fig:PMT2}b shows the clearances of a typical installation.

\subsection{Alignment and Survey}   

As noted previously, normal construction techniques produce acrylic and stainless steel structures varying in size by  several mm.
It is also difficult to precisely position large objects such as the OAV with overhead cranes.
The trade-off between cost and allowed tolerances was best optimized by adopting an AD  design 
which allowed up to centimeter deviations in the acrylic vessel height and port locations. 
Detector simulations~\cite{Bryce} show that variations in component size or offset result in only minor variations in detection efficiency.  
Thus, in nearly all cases, engineering tolerances on detector design were stricter than design requirements
based on the desired similarity of detector response.

Survey of the ADs with a Leica TPS1200 Total Work Station~\cite{Leica} occurred at many stages of the assembly process to both record the 
parameters of the as-built detectors and to align the AV calibration ports with the SSV lid. Good alignment of the calibration ports was 
necessary for reliable deployment of the calibration sources into the inner volumes of the detector.
The Leica had a nominal resolution of $\pm~0.2$mm and a repeatability of $\pm~1.0$mm.

Specialized tooling 
as shown in Fig.~\ref{fig:Survey}d was designed to aid in positioning the acrylic vessels.
The OAV clamp-downs were designed oversize to allow the OAV to be shifted, rotated and centered within the SSV.
The IAV was mounted on four teflon pucks plus shims which could adjust the tilt of the IAV within the OAV to align the calibration ports.
The calibration tubes were made from convoluted teflon cylinders which accommodate misalignments of a few  cm or less.
In addition each calibration tube had a sliding O-ring connection to the SSV lid which could accommodate height
differences of 2-5 cm.

An initial survey of the SSV lid established a local coordinate system centered on the central overflow calibration flange and referenced
to bolt holes on the perimeter of the SSV lid. 
Thus measurements taken with the SSV lid off could be referenced to the nominal coordinate axes. 
Fig.~\ref{fig:Survey} shows some typical survey procedures.
The sequence of survey measurements was:

 \begin{figure}\hfil
\includegraphics[clip=true, trim=0mm  10mm 0mm 10mm,width=5.0in]{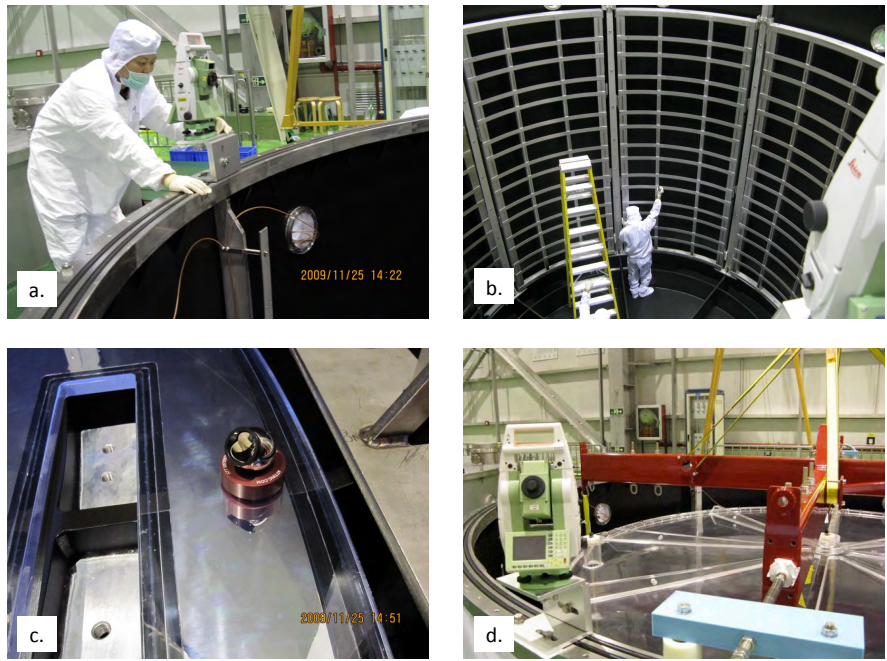}\hfil%
\caption{(a) Setting up Total Survey Station on SSV flange. (b) Measuring radius of PMT ladders.
 (c) Survey of height of lower reflector. (d) Adjusting the position of the OAV to align the central port with the SSV. } 
\label{fig:Survey}
\end{figure} 

\begin{itemize} \itemsep1pt
\item Establish the circumference of the SSV lid and calibration port locations.
\item After the SSV lid was removed, measure standard points in the interior of the SSV.
\item Install and align the PMT ladder rails. Install the PMT ladders and measure standard points on each ladder frame.
\item Measure the height of the SSV ribs and and the center hub.
\item Install and survey lower 5 cm monitoring PMTS.
\item Install the lower reflector and measure standard points on the reflector surface.
\item Install the OAV into the AD while aligning the central calibration port with the SSV centerline. Rotate the OAV to best align the outer ports.
\item Secure the OAV to the SSV, remove the OAV lid,  and measure a standard set of points on the OAV floor, ribs, and cylinder wall.
\item Install and shim IAV teflon support pucks on the bottom of the OAV, temporarily install the IAV and measure the position of the ports.
Remove the IAV and adjust the puck heights to bring the IAV ports into the desired position.
Verify the IAV port positions, readjust pucks as necessary.
\item Install the OAV lid and re-survey the port position, rib heights, and other points on the OAV lid.
\item Install the upper reflector and survey standard points on the reflector surface.
\item Install and survey upper monitoring PMTs on underside of the SSV lid.
\item Install the SSV lid and LS overflow tank and measure standard points on the overflow tanks.
Measure the position of the off axis cameras.
\end{itemize}

\begin{figure}\hfil
\includegraphics[clip=true, trim=0mm 0mm 0mm 0mm,width=6.0in]{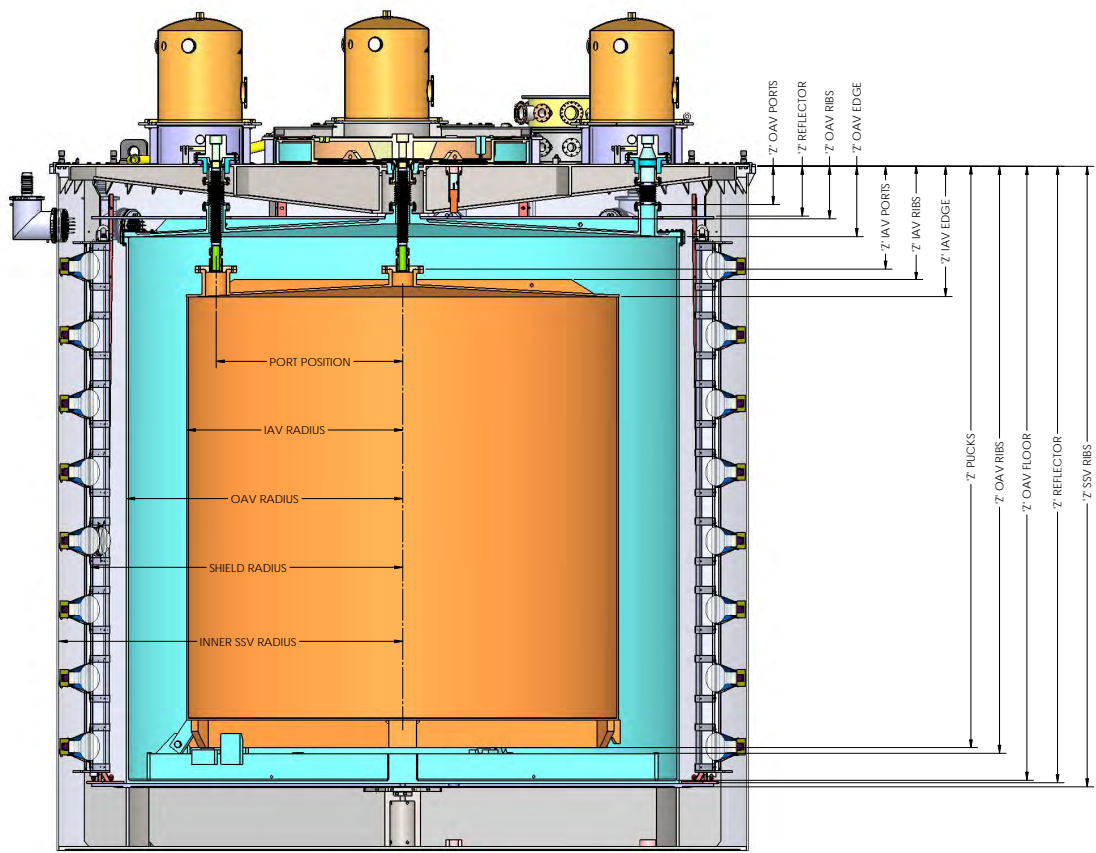}\hfil%
\caption{Some of the survey data points available for each AD. 
Measurements at multiple locations sampled variations in the shape of the reflector, SSV, IAV and OAV.} 
\label{fig:Survey_data}
\end{figure}

Fig.~\ref{fig:Survey_data} shows some of the survey data available for each AD. Height data was taken at multiple
locations on the SSV ribs, lower reflector, internal and external OAV ribs, IAV external ribs, upper reflector and acrylic vessel
ports. Similarly, data was also obtained on the radial positions of the IAV, OAV, SSV, and PMT ladder walls.

\begin{figure}\hfil
\includegraphics[clip=true, trim=0mm 0mm 0mm 00mm,width=5.5in]{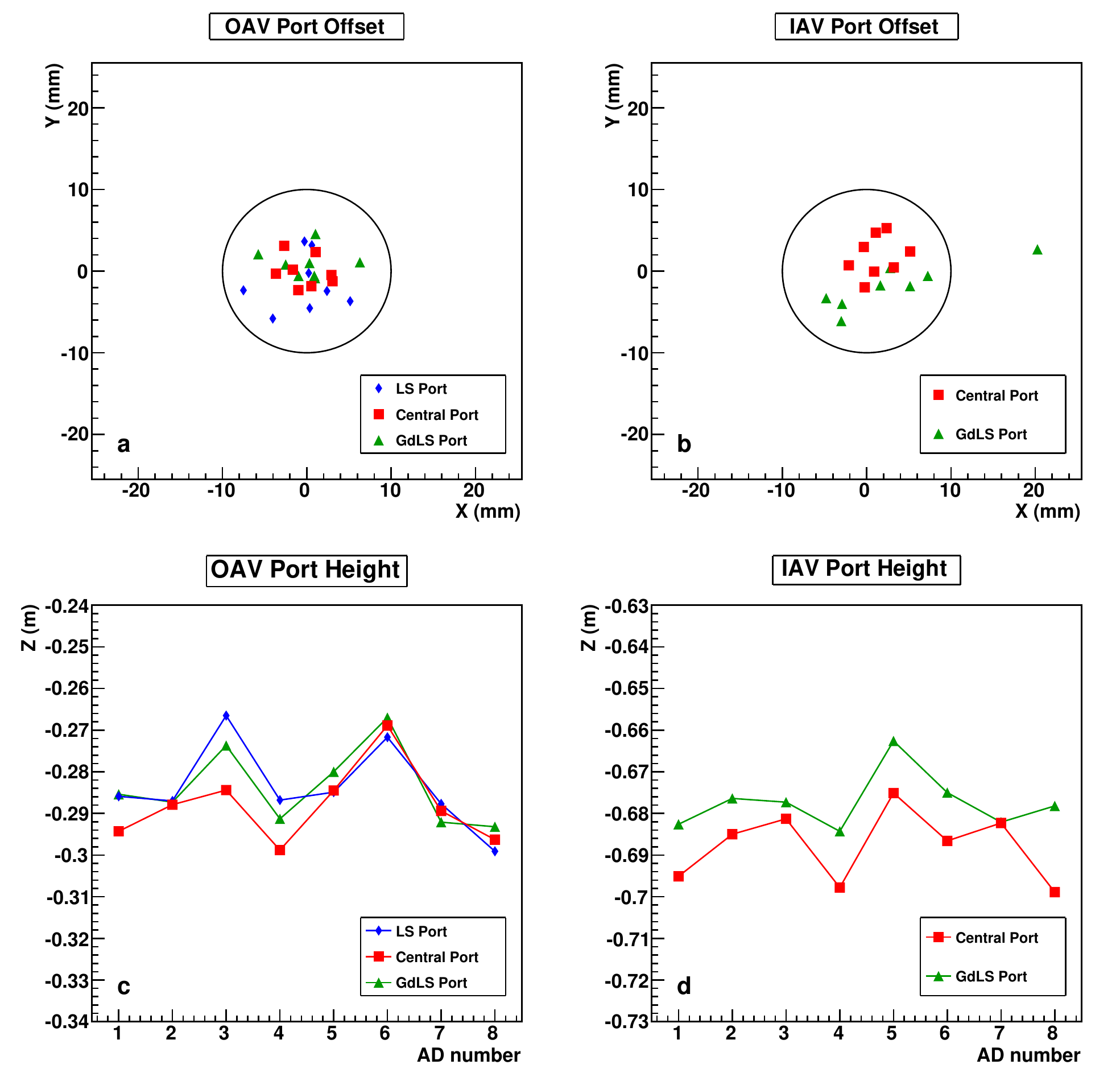}\hfil%
\caption{Measured XY locations of the OAV and IAV calibration ports relative to the SSV port are shown in a and b.  A 1 cm radius circle is shown for scale.
The z positions relative to the SSV lid are shown in c and d have a larger spread.} 
\label{fig:Port_locations}
\end{figure}

During  AD assembly
particular care was taken to align the IAV and OAV calibration ports with the openings in the SSV lid
to minimize interference with the calibration sources which are lowered through these ports. The measured XY offsets
are shown in Fig.~\ref{fig:Port_locations}a and b. All but one of the ports are well within 1 cm of the nominal location.  
A construction error in the first AD caused the 2 cm offset in the IAV off-axis port. 
The z locations of the AD ports are 
shown in  Fig.~\ref{fig:Port_locations}c and d. In many cases the central port is lower 
than the off-axis ports indicating that the central lid
has sagged in the center without the presence of the final liquids.

\begin{figure}\hfil
\includegraphics[clip=true, trim=0mm 0mm 0mm 0mm,width=5.5in]{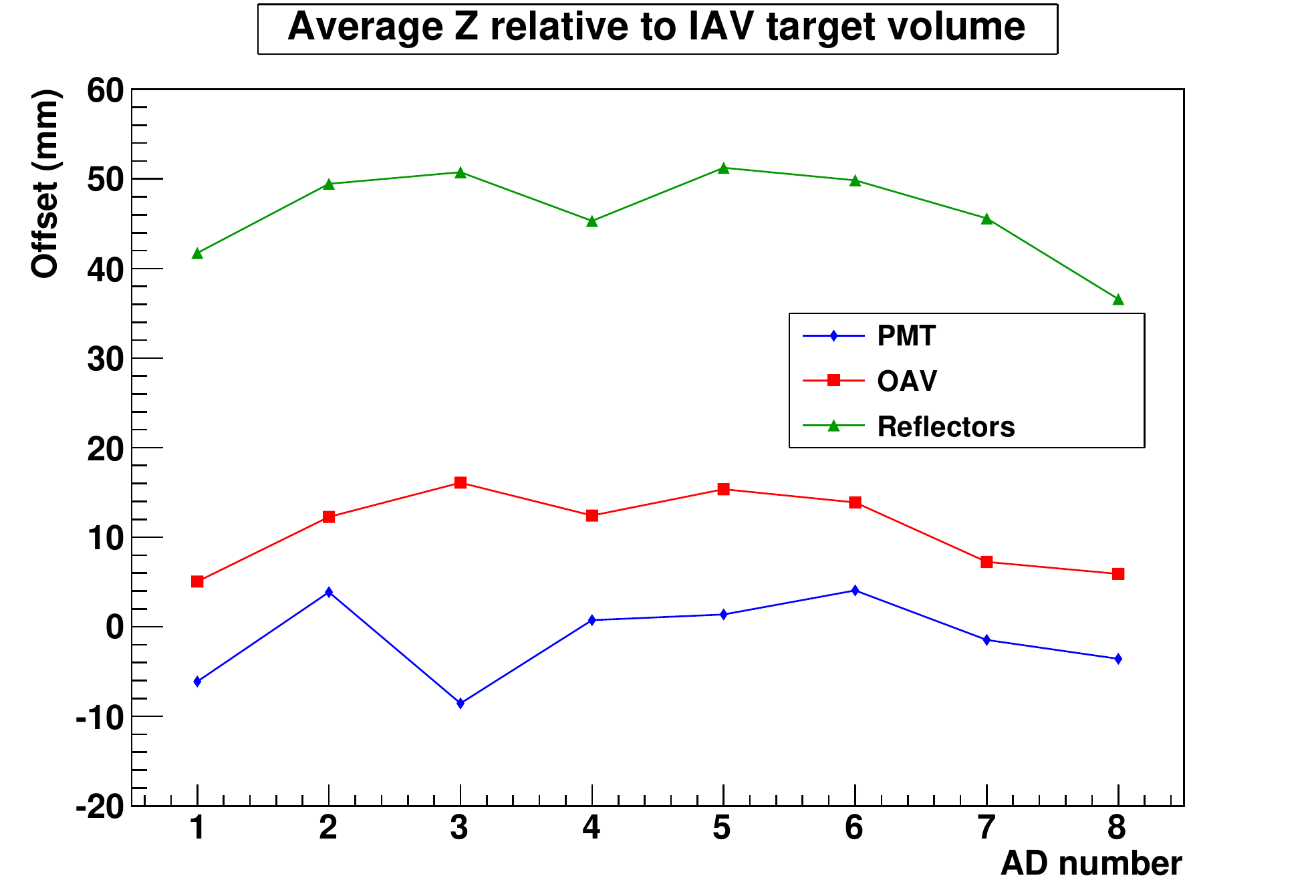}\hfil%
\caption{Average z position of detector elements relative to the center of the GdLS target mass.} 
\label{fig:ZZ}
\end{figure}

The survey data can also be used to characterize  parameters which may alter the light collection 
efficiency between differing ADs. Fig.~\ref{fig:ZZ} compares how well various detector elements are centered 
vertically on the center of the visible GdLS target mass volume contained in the IAV. The average GdLS z is calculated from a combination of survey 
and construction data with a correction for the volume in the cone shaped IAV lid. Although the PMTs are centered on the GdLS volume within a cm, the 
center of the OAV is systematically displaced from the IAV center by $\approx 1$~cm.  
By design, the difference between the OAV top and bottom rib heights and the conical lid
cause the reflectors to be placed asymmetrically causing the observed 4.5 cm offset between the center of the reflectors and the center of the IAV.
The RMS spread observed about these central
values is 0.5 cm. 

Other dimensions that could impact the light collection efficiency are the average radii of the PMTs and 
the space between the upper and lower reflectors.
The position of the  PMT photocathodes are  measured from the highest point of the glass bulb to the surface of the radial shield.
Although the spread in the measured PMT ladder radii can be over 3 cm  within a single AD, the average AD radii  are consistent with
a standard deviation of 0.2 cm.  Seven of the eight ADs  have similarly consistent distances between the two reflectors ($\sigma = 0.2$ cm), however in
the last AD this distance was 1.3 cm smaller.  In general, most of the physical dimensions of the ADs were within 0.5\% of each other and often 
within 0.05\%.  Ongoing studies of the AD energy calibrations have yet to find any obvious correlation between the measured geometrical differences
between detectors and the detector energy response.

\subsection{Leak checking}     

Each AD has nearly 1000 O-rings and gaskets separating  gas and
liquids inside the AD and  water pool. Any significant leak 
would have severe consequences.  Water leaks into a dry-box containing the AD PMT cable interconnections could short an
 entire ladder of PMTs (1/8 of total).  Water leaks into the ACUs could short out 
 the motor drives and drip into the Gd-LS. Leaks between the LS and GdLS 
 volumes would introduce systematic errors in the calculation of the target mass 
 and energy calibration. After the AD is fully assembled it would be difficult to locate 
 the source of a leak and would require a major intervention (draining the water pool) 
 to repair. All of these factors made it  important to identify leaks as early and reliably as possible.
 
Although the obvious goal was to have no measurable leaks, it was necessary to set practical leak rate goals that were
achievable and were based on realistic estimates of the impact of small leaks. For example a 3 liter leak between the GdLS
volume and the LS volume would change the overall target mass by 0.01\%, a small amount compared to the target mass uncertainty of
0.03\% and the  $\sim$~0.2\% relative systematic uncertainty in the AD detector efficiency. 
Assuming a five year operational lifetime,  a 3 liter leak over 5 years would be Q =  $18.6 \cdot 10^{-6}$ cc/sec.
or 1.5 cc/day.  
It would be very difficult and take a long time to identify such a leak reliably. In addition the liquids 
inside the AD were not  available until very late in the assembly process. 
To address these issues,
leak checking was done with gases
 (air, argon or freon 134a) and the results scaled to the appropriate liquid.

To compare leak rates of the different gas and fluids it is useful to start with  the Hagen-Poiseuille
 equation which relates the expected  leak rate for an pinhole leak  of diameter d and length L.
\begin{equation}\label{eqn:HP}
Q = \pi \cdot d^4 \cdot  \Delta P \:  /  \: (128  \cdot \mu \: \cdot \: L)
\end{equation}
Q is  the volume flow,  $\Delta$P is the pressure difference, 
 and $\mu$ is the dynamical viscosity of the liquid or gas.
Note that the leak rate is proportional to the pressure difference across the pinhole.
Leak rates scale inversely with the dynamical viscosities, $\mu$, of the gas or liquid
shown in Table~1.
 For fixed d, $\Delta$P, and L an air leak is
  55 times larger than a water leak or  
 884 (3097) times larger than a LS (MO) leak  when measured near atmospheric pressure.

\begin{table}[htdp]
\caption{Dynamical Viscosity of relevant  liquids and test gases. Values for GdLS and LS 
are estimated from the linear alkyl-benzene (LAB) solvent.}
 \begin{center}
 \begin{tabular}{|c|c|c|c|}
\hline
   Liquids              &Pa-sec                          & Gases                        &    Pa-sec \\
   \hline
    water                 &   $1 \cdot 10^{-3}$    & nitrogen                      & $19 \cdot 10^{-6}$ \\
\hline
LAB                       & $15 \cdot 10^{-3}$    & air                                 &   $18 \cdot 10^{-6}$ \\
\hline
mineral oil           & $52 \cdot 10^{-3}$    & argon                            & $22.5 \cdot 10^{-6}$  \\
\hline
                              &                                      & freon134a                    & $13.9\cdot 10^{-6}$   \\
\hline
\end{tabular}
\end{center}
\label{gases}
\end{table}

There is an additional consideration when comparing gas leaks to liquid leaks through the same opening.
 Unlike liquids, which typically do not change volume with pressure, the gas density is proportional to the pressure.  
 The volume leak rate in standard cc/sec (at standard conditions, 1 atm., $60^{\circ}$F) is given by~\cite{Jolly}:  
  
  \begin{equation}\label{eqn:HP2} 
  Q_{gas} = \pi \cdot d^4 \cdot \Delta P \cdot P^{\prime}_a/(128  \cdot  \mu \cdot L) 
   \end{equation}

The term $P^{\prime}_a$ is defined as $(P_i+P_o)/(2 \cdot  P_o)$ which is the average pressure (absolute) of the gas inside and outside the leaking vessel.  When $P_i$ and  $P_o$ are $\sim$1 atm, the equation is identical to the Hagen-Poiseuille equation.
In general any real leak is unlikely to be a perfect capillary in shape or have a constant 
cross-section along its length.  The $\pi \cdot d^4$/L term can be replaced by a constant K (the same 
for both gas and liquid), which is determined from the gas leakage measurements and used to predict the liquid leak rate. 
The relative gas and liquid leak rates through the same opening then becomes 

\begin{eqnarray}
\frac{Q_{\rm gas}}{Q_{\rm liquid}}&=   P^{\prime}_a  \cdot&
\left (\frac{\mu_{\rm liquid}}{\mu_{\rm gas}}\right ) 
\left (\frac{ \Delta P_{\rm gas}}{ \Delta P_{\rm liquid}}\right)
\label{eq:gasratio}
\end{eqnarray}
where $P^{\prime}_a$ is in absolute atm.  Eq.~\ref{eq:gasratio} is used to scale a measured leak rate using a test gas at specified  
differential and absolute pressures to the corresponding leak rate for the liquid of interest 
under the maximum  differential pressure expected under normal operating conditions.

Leak test goals varied with the amount of the allowed liquid leak over 5 years (typically 2-5 l), 
the differential pressure (0.12 - 3.0 m of water equivalent),
and the viscosity of the liquids.  
The most stringent criteria were applied to joints exposed to the water pool which had the 
lowest viscosity liquid (water) at the highest differential pressure (2.5 - 3.0 m of water) since the interior of the AD is
at atmospheric pressure. More relaxed criteria were used for joints between the various acrylic components
since the pressure differentials were limited to < 15 cm of water by design and 
because of the high viscosity  of LS and MO.   

\begin{figure}\hfil
\includegraphics[clip=true, trim=20mm  70mm 20mm 65mm,width=6.0in]{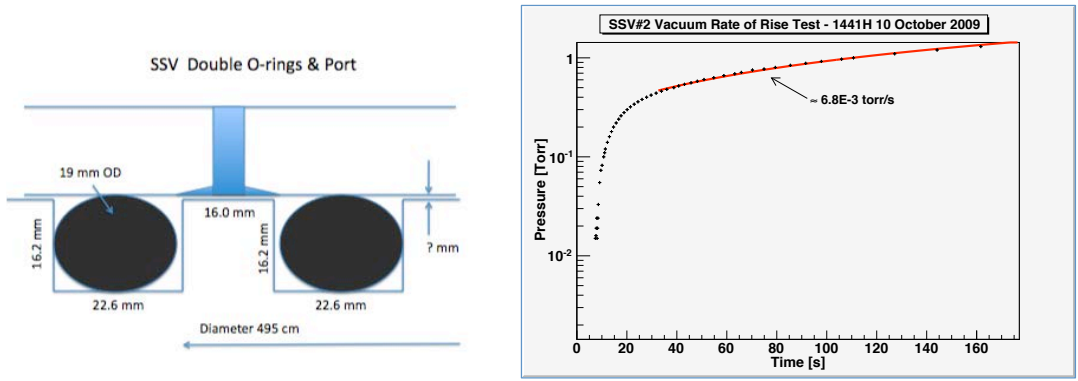}\hfil%
\caption{(left) Drawing of the double O-ring design for the SSV lid.   (right)
 Typical vacuum rate of rise measurement.  }\label{fig:leak_SSV}
\end{figure}

Several methods of  differing sensitivity were used for leak checking. Where possible, a 
vacuum pump-down followed by a vacuum rate of rise measurement was used since this had the best 
sensitivity and tested the entire surface area or volume. All bellows used on the lid of the AD or 
used to connect the AD to the outside of the  water pool were pre-tested by this method.  
Vacuum tests were also made on  the 
double O-rings used in most of the overflow tank covers, ACUs, and port covers. For example the pump-out 
port between the O-rings on the SSV lid, shown in Fig.~\ref{fig:leak_SSV},  was used to evacuate the air  
between the O-rings  until the pressure fell to $\approx 1 $mTorr. The vacuum pump was then valved off and the
 vacuum rate of rise was  measured and plotted in Fig.~\ref{fig:leak_SSV}.  Using the known volume under vacuum, the leak rate 
 could be calculated. Leak rate goals of $\approx 1.5 \cdot 10^{-3}$ cc/sec were
 achievable despite significant out-gassing of the O-rings. 

Since the acrylic vessels had been visually inspected for flaws as part of the acceptance tests,
no specific leak checks of the acrylic surfaces were made. Only the O-ring seals to the calibration tubes or OAV lid
were tested. The large acrylic vessels were fragile and could not withstand significant pressure
 differentials.  To test the connections of the calibration tubes to 
 an IAV or OAV the vessel was first filled with > 75\% argon gas. The vessel was then 
 carefully pressurized to $\approx 15$ cm of water equivalent. An argon leak checker~\cite{Matheson}
 was used to sniff for argon leaks at the O-ring joints. 
 The argon sniffers could easily identify leaks  $> 10^{-2}$ cc/sec. 

When greater sensitivity was required (as in leak checks of the 
various bellows  after installation), freon134a was used to fill the  volume being tested to a few psi.  A freon sniffer~\cite{Accuprobe,TPI}
was used to inspect all joints. These devices were sensitive to leaks smaller than $\approx 1.5 \cdot 10^{-3}$ cc/sec
and were often used during the final installation of an AD into the experimental hall.
A combination of these methods and some specialized tests described later were used to leak check the AD during assemble and installation.
Both argon and freon sniffers were calibrated using calibrated leaks~\cite{Laco}. All sniffers could detect leaks ten times smaller 
than required.  

\begin{table}[!h]
\caption{Summary of the different leakage tests performed on each AD during assembly and installation.
The maximum differential pressure ($\Delta~p_l$) expected under nominal conditions is given in cm of water equivalent.
The typical differential pressure ($\Delta~p_g$) of the test gas is given in cm of water equivalent or in absolute atmospheres.
Multiple test methods were utilized in some cases.}

 \begin{center}
   \begin{tabular}{b{4cm} | c c c c c c c }
\hline
        & \begin{sideways}5 year leak goal\end{sideways} & \begin{sideways}\# per AD\end{sideways} & \begin{sideways}$\Delta p_{l}$ [cm]\end{sideways} & \begin{sideways}Test Method\end{sideways} & \begin{sideways}Test Gas \end{sideways}& \begin{sideways}Test $\Delta p_{g}$\end{sideways} & \begin{sideways}Test Goal [cc/s] \end{sideways} \\
\hline\hline
\bf{GdLS to LS}   &  & &   &  &  & [cm] &  \\ 
 
\hline
Calibration tubes  &   3 l& 2& 15        & sniffer       & Ar   & 15  &   $1.3 \cdot 10^{-2}$ \\
IAV  body          &  3 l & 1& 15    &freon fraction     &freon    &12          & $1.7\cdot 10^{-2}$\\
\hline\hline
\bf{LS to MO}    &  & &   &  &  &  &  \\ 
\hline
Calibration tubes   &   5 l & 3& 15        & sniffer            &Ar   & 15 cm              &   $2.1 \cdot 10^{-2}$ \\
OAV lid O-rings     &   5 l & 1& 15        & pressure drop          &Ar   & 0.3 atm            &   1.7 \\
OAV body            &   5 l  & 1& 15                            & freon fraction          &freon    &12 cm                & $2.8\cdot 10^{-2}$ \\
\hline\hline
\bf{MO  to N$_2$}   &  & &   &  &  & [atm] &  \\
\hline
\raggedright PMT cable feedthroughs & 2 l  & 240 & 50        & vacuum rise    &air  & 1  &  $3.8\cdot 10^{-1}$ \\
                       &  & &        & sniffer            &Ar   &     0.3      &   $2.4\cdot 10^{-1}$ \\
AD cameras    & 0.05 l & 2&330        & sniffer       &freon  & 0.3  &   $1.5\cdot 10^{-3}$ \\
MO clarity   &  0.12 l & 1& 15        & pressure drop           &freon  & 0.14   &   $3.6\cdot 10^{-2}$ \\
\hline\hline
\bf{Water to MO or N$_2$}    &  & &   &  &  & [atm] &  \\
\hline
SSV lid  O-rings      & 2  & 1& 250       & vacuum rise            &air   & 1             &   $1.5\cdot 10^{-3}$ \\
Overflow tank  O-rings & 2 & 3 & 250        & vacuum rise         &air   & 1             &   $1.5\cdot 10^{-3}$ \\
MO fill port           &  2 & 2& 250   & vacuum rise         &air   & 1           &   $1.5\cdot 10^{-3}$ \\
MO clarity             & 1.6  & 1& 250  & pressure drop        &freon  & 0.14         &   $4.\cdot 10^{-4}$ \\
ACU                    &  2 & 3& 250  & vacuum rise          &air   & 1          &   $1.5\cdot 10^{-3}$\\
\raggedright Electrical, gas, PMT cable \& supply bellows     & 2& 30  & 250 & vacuum rise          &air   &1   &   $1.5\cdot 10^{-3}$ \\
Bellow connections     & 2  &60 & 250   & vacuum rise          &air  & 1   &   $1.5\cdot 10^{-3}$ \\
                       &   & &        & sniffer         &freon  & 0.3          &   $6.5\cdot 10^{-3}$ \\
SSV drain plug         &   2& 1& 250        & vacuum rise          &air   &1   &   $1.5\cdot 10^{-3}$\\
\hline
 \end{tabular}
 \end{center}
 \label{tab:allleak}
\end{table}

\begin{figure}\hfil
\includegraphics[clip=true, trim=20mm  50mm 20mm 50mm,width=5.0in]{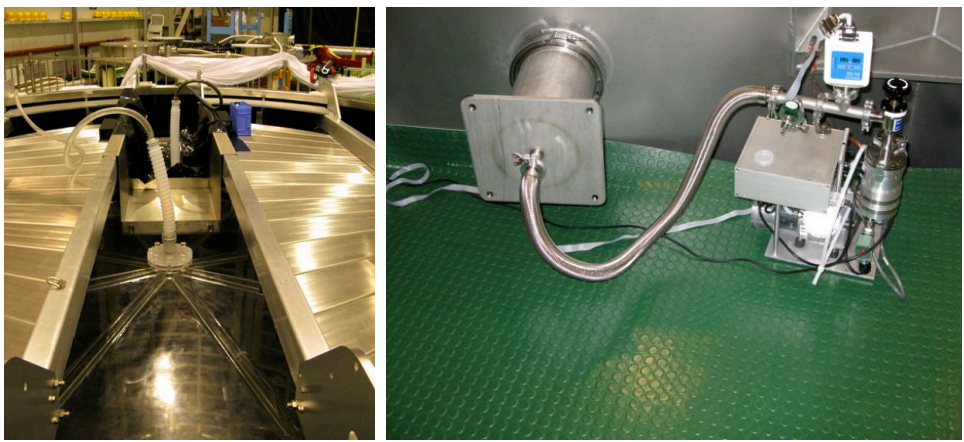}\hfil%
\caption{(left) The IAV was filled with more than 75\% argon at 12cm of water equivalent pressure for a leak test.   (right)
Leak test of the PMT cable flange with the Top hat.
 }\label{fig:Leak}
\end{figure}

Leak checks were performed  at specific stages of the assembly process. 
After the IAV vessel was positioned in the OAV, the port covers capping the calibration tubes were
tested by filling the IAV with argon at 15 cm of water pressure and sniffing the O-ring joints as shown in  Fig.~\ref{fig:Leak}
After the OAV lid was installed the double O-rings were pressure tested at 5 psi. The lower parts of the OAV calibration tube and the
connections to the OAV flanges were also tested with argon. No significant leaks were found in any of the 8 ADs tested.
The numerous leak tests  performed on each AD are summarized in Table~\ref{tab:allleak}.

As discussed in Section 2.4, the PMT  cables from each PMT ladder exit the SSV volume through ~30 custom plugs mounted in a 
vacuum flange assembly  mounted onto the side of the SSV. Each plug has 
four O-rings, two sealing the plug to the hole in the dry-box flange and two sealing the cable to the plug. 
A bell shaped test chamber (top hat) shown in Fig.~\ref{fig:Leak}     was 
designed to cover a fully populated dry box flange enabling a vacuum test of all the plugs at one time. 
About half of the dry-box flanges
failed the initial vacuum test. Failed dry-boxes were further tested by filling the test chamber with argon at 5 psi and using an 
argon sniffer to locate leaks. Although some missing or damaged O-rings were discovered and repaired,
 the most common problem was
a scratch or hole in the outer jacket of one of the coaxial cables.  Air could leak through the hole and travel along the cable 
ground braid and escape through the cable connector inside the test chamber. The size of the leak varied with the size and location
of the hole. Those cables with leaks  $\geq 1.5 \cdot 10^{-2}$ cc/sec (air) were replaced. In all cases the final repaired
flange passed the vacuum test.

Prior to installation of the SSV lid, all gas and electrical interface flanges were tested by the vacuum technique. The
gas and electrical bellows containing the cover gas, lid sensor, and ACU lines were pretested using the vacuum technique.
After installation the joints between the bellows and the SSV lid, interface boxes, or ACUs were retested by the vacuum technique when possible.
Other joints were tested by sniffing the joints with a freon sniffer after the inner gas volume was filled with freon at 5 psi.  
As a final precaution, the  PMT and electrical gas volumes were pressurized to 0.5 m of water equivalent prior to the filling of the
water pool so that leaks would cause visible bubbles when the water covered each joint.

After the installation of the SSV lid,  the upper end of the calibration tubes were connected  to the lid or  overflow tanks
completing all connections to the IAV and OAV.  A final specialized test was then made of the sliding O-ring joints in the 
calibration tubes  and the entire IAV and OAV volumes. This was done by making  separate gas connections to the IAV, OAV, and 
the space between the OAV and SSV (SSV-OAV). 
The OAV plus OAV calibration tube volume completely surrounds the IAV plus IAV 
calibration tube volume and is surrounded in turn by the SSV volume. The OAV volume is then filled with $\ge 75$\% freon134a  at
$\approx 12 $~cm of water pressure for several days. The other two volumes are maintained near atmospheric pressure during the test.
Any leak would allow freon to flow from the higher pressure OAV volume into one of the other volumes. The freon content of the
other volumes was measured before, during, and after the test~\cite{BigDipper} and checked with the freon sniffers.
Assuming that a leak would be diluted by the entire IAV or SSV-OAV  volume (worst case) daily increases of
>105  and >53 ppm would be observed in the IAV or SSV-OAV volume for leaks larger than the goal of $\approx 1.7 \cdot 10^{-2}$ cc/sec. 
This test identified one AD in which the 3m calibration tube assembly was found to be too short to seal reliably. A  new longer part 
was fabricated, installed and retested  successfully.

\section{AD Installation}

Completed ADs were transported to the Liquid Scintillator hall for filling.
Pairs of ADs were filled within one week of each other. Details of the filling process can be found in 
reference~\cite{Fill}.
Filled ADs were then transported to experimental halls with the AGV and placed on temporary stands by the water pools. 
The transportation of the first AD was monitored live by the highest of the two monitoring cameras. No internal movement  of any of the acrylic
vessels or calibration tubes was detected. 
The ADs were then prepared for installation and lifted into the water pool.  
Access platforms were installed to enable the connection of gas and electric lines
from the AD to the gas or electronics rooms.
Eight bundles of  PMT  cables were connected to the 35 m long cables from the electronics room.
Automated calibration units (ACUs) were mounted over each access port .
All lid electrical cables were routed to an electrical distribution box and through a single
large diameter bellows to the edge of the water pool and then on to the electronics room.
Final electrical and leak checks were made before 
filling the water pool and sealing the top with a light-tight pool cover.
At this point installation was complete and commissioning of the detectors could begin.

\subsection{Experimental Halls}  

\begin{figure}\hfil
\includegraphics[clip=true, trim=35mm  45mm 45mm 45mm,width=5.0in]{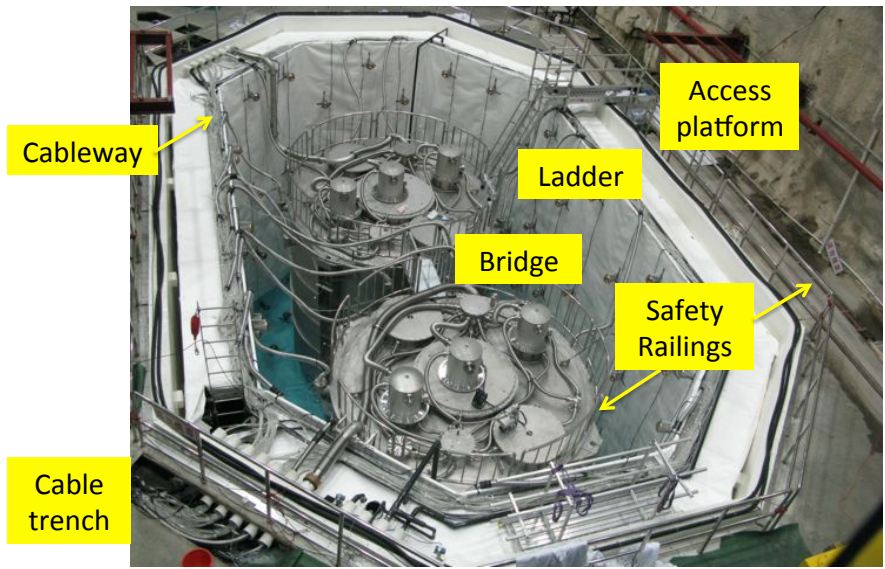}\hfil%
\caption{A near hall nearing completion. 
The access platform and ladder provided access to the 
top of the detector  where final gas and electrical connections were made. 
Railings around the AD and pool protect workers
from the 7-10m drop to the pool floor. 
The cableway running around the pool and the cable trenches outside the pool are also shown.}\label{fig:Int_EH3}
\end{figure}

The ADs were installed in  water pools instrumented with PMTs in the underground experimental halls  
as shown in  Fig.~\ref{fig:Int_EH3}.
The far hall contained a 16m $\times$ 16m $\times$ 10m (deep) water pool holding four ADs. 
The near hall water pools were 10m $\times$ 16m  and held two ADs. 
The water provided both passive shielding  of the radioactivity emitted by the cavern rock and concrete and active detection
of cosmic rays impinging on the ADs.
Next to the water pools, gas and electronic rooms housed the data acquisition, control and monitoring electronics,
and gas supply racks. PMT  cables  and gas lines were run in trenches to the edge of the water pool and then in
cable ways mounted above the water level.  Stainless  steel bellows protected the cables and gas lines from the water.

Custom built platforms provided temporary access to the top of each AD  as seen in   Fig.~\ref{fig:Int_EH3}.  
Railings  were mounted on the ADs for safety since the AD lids  were 7m above the bottom of the pool.
A specialized platform was built to install the PMT cables. This platform hung over the side of the AD allowing
workers to access the dry box connections that were about 0.5 m below the lid. 
        
\subsection{Installation Sequence }

\begin{figure}\hfil

\includegraphics[clip=true, trim=35mm  10mm 35mm 10mm,width=6.0in]{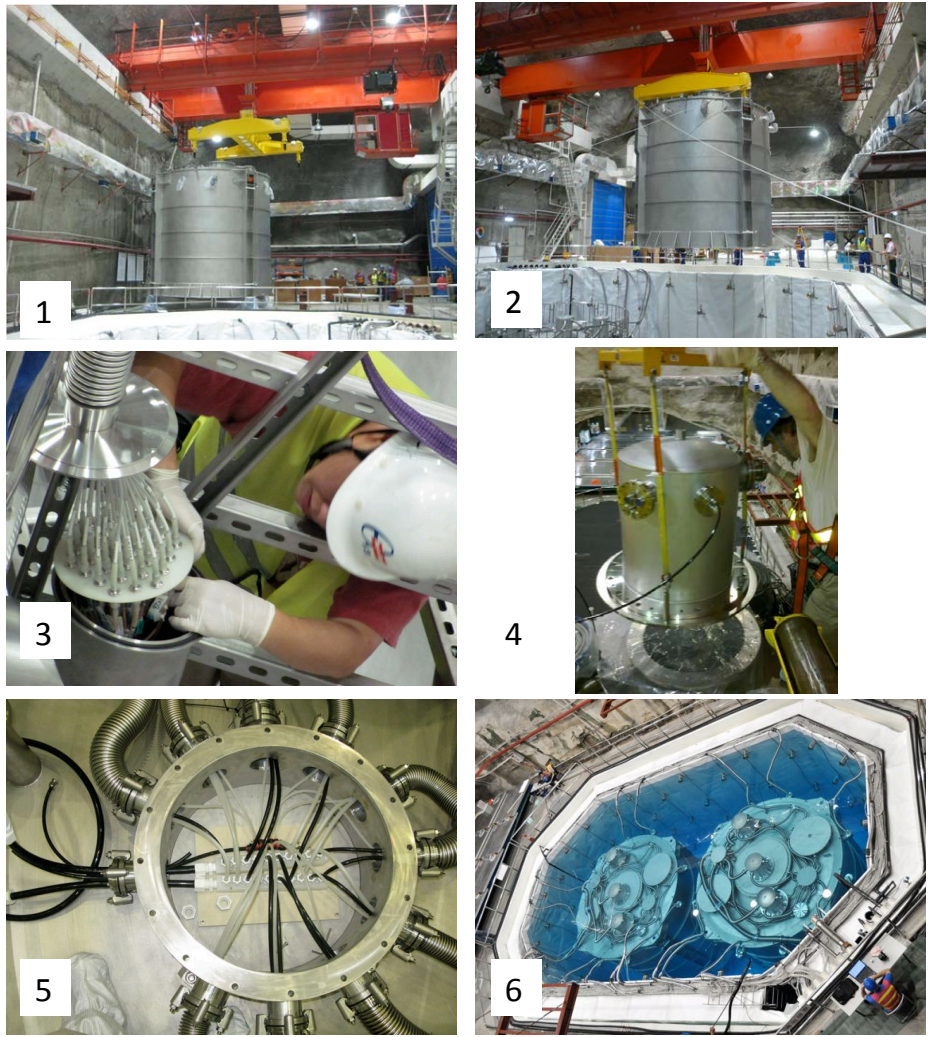}
\caption{Installation sequence. 
1) AD is prepared for lift into pool. 
2) Lift into pool. 
3) PMT cables connected
4) ACUs installed.  
5) Gas lines installed. 
6) Water pool filled.}\label{fig:InstallB}
\end{figure}

The typical installation sequence is shown as a series of photographs in Fig.~\ref{fig:InstallB}. 

\begin{enumerate} \itemsep1pt
\item An AD is delivered from the LS hall on the automated guided vehicle (AGV) and placed on temporary stands next to the water pool.
The move dislodges gas bubbles under the OAV and SSV lids lowering the  liquid levels in the overflow tanks.
The level change is measured by the AD levels monitors. The fluid levels are then manually topped up to the desired levels.
Drybox elbows were mounted over the PMT feedthrough ports. Muon pool calibration LEDs were mounted on the side of the AD.
Finally the heavy lifting fixture and positioning cameras were mounted in preparation for the lift into the pool.
\item
The AD is positioned on its support stands inside the pool and clamped down. The stand and clamps are electrically isolated 
from the AD.  Access platforms are rigged into position and work on top of the detector begins.
The AD is connected to the cover gas system \cite{Gas}. 
\item
Eight PMT cable bundles are routed to the electronics room per AD.
A bundle of PMT cables for one PMT ladder is threaded through a 3 inch bellows.
The upper end of the bellows is secured to the cableway at the periphery of the pool. 
The remaining cable is routed to the electronics room via the cableways and trenches shown in  Fig.~\ref{fig:Int_EH3}.
The AD end of the bellows is lifted into the pool and the PMT cables are connected inside the dry box elbow
as seen in Fig.~\ref{fig:InstallB}. The bellows assembly is then fastened to the elbow by a vacuum style flange.
\item
ACUs are installed over the calibration ports. All of the lid sensors and ACU controls are routed to the electronics room
via an electrical distribution box and a 6 inch diameter bellows that extends to the edge of the pool.
\item 
Gas connections to the ACUs are finalized and leak checked.  
The PMT, electrical and gas line bellows are independent
gas volumes distinct from the gas in the overflow tanks and ACUs and are leak checked separately.
\item
The lid is wiped clean and the water pool is filled. 
The AD is monitored for any signs of leaks. 
After filling the pool cover is installed and AD commissioning starts.
\end{enumerate}
    
\subsection{Final installation checks }     

Multiple leak checks of the final gas and electrical connections were necessary during installation of an AD into the water pool. 
Functionality tests of the lid sensor and PMT control cables were made as soon as possible  after the ACU installation.
The PMT  cables were checked for continuity shortly after connection. PMT testing started after  the ACU installations were complete
and the AD interior was completely dark.

There are four separate gas volumes on each AD. The cover gas fills the spaces above the overflow tanks and the ACUs. 
The cover gas lines run inside bellows connected to the gas distribution box and onto the surface. The electrical cables run through
a similar set of bellows. Each drybox and bellows were separate gas volumes underwater but were run in parallel with manifolds in the
trenches.  Although the integrity of all of the bellows had been vacuum tested before installation the interconnections between 
the bellows had to be tested. Where possible this was done by a vacuum test method. 
However, the final test~\cite{Gas} was fill the volume with freon
at a few psi and to use a freon sniffer to inspect the connections and O-rings.

\subsection{Position Survey }       

A final survey was made of the monuments on the SSV lid with respect to monuments on the experimental hall wall. 
The wall coordinate system had been previously connected with the position of the reactor cores. 
Together the surveys measured the distance from each detector to each reactor with a precision better than 18 mm.

\section{Lessons Learned}

The original AD design proved to be robust and was changed in only a few areas. As detailed in section 2.4
the PMT ladders  were not perfect cylindrical sections as designed, necessitating  the addition of a position adjustment
clamp to allow the PMTs to be moved radially inward as needed.

In the original design the long PMT cables from the electronics room to the AD were to run through the water pool
before entering  the dry box elbow to connect to cables from the PMTs. Although the O-ring seals around the cables
were reliable and worked well,  leaks  through the cable jackets could propagate through the cable braid inside the
cable jacket into the dry box. The size of the leak depended on the size of the hole in the cable jacket and how far from the 
dry box it was.  Cable leaks were not infrequent and were quite difficult to identify. It soon became apparent that leak checks
would not only take much longer than practicable but that damage to the cables during installation in the water pool could result 
in further leaks.

The cable routing plan was modified to include a  protective 65 mm bellows that ran from above the pool waterline 
to a vacuum style flange on the dry box elbow. A custom compression plug~\cite{Gas} provided a gas seal on the upper end of the bellows were 
the cables exited into the cableway. The dry box and bellows assembly was leak checked by filling with freon at a few psi 
and sniffing all joints with a freon sniffer. Installation and leak checking of the 8 PMT cable bundles
were greatly simplified by this design change and took less than two days per AD.  

The OAV lids were machined from two large sheets of cast acrylic. The first lid had some small areas of crazing at the joint between the sheets 
after the OAVs arrived at Daya Bay.  The lid was replaced. Although no definitive cause of the crazing was identified, the shipping container packaging
was changed to add insulation to the top of the shipping container to reduce the large day/night temperature swings observed 
when the AD was in the shipping yard. No further problems were experienced with the AD lids.

 The second pair of IAVs were built incorrectly with the top lid rotated with respect to the bottom by 45${^{\circ}}$.  A modification to the last
 pair of OAVs allowed these IAVs to be used in the last AD pair. One of these  IAVs also had 
 crazing in the acrylic wall  and was damaged during a repair attempt at the Daya Bay
 site. A ninth IAV was built to replace the damaged IAV. 
 
 During tests of the manual calibration system on AD8 the cable supporting the source arm failed, dropping the 20 kg load onto the floor of the IAV.
 The resulting 20 cm crack was repaired by removing the IAV from the SSV/OAV  and cutting out the damaged section. A new custom piece was
 bonded into the hole and the entire IAV was re-annealed.  The entire AD was then re-assembled.  Despite its history AD8 appears to function
 similarly to the other ADs.

\section{Summary}

The design, planning, construction, assembly and installation of the eight Daya Bay antineutrino detectors took over four years and the efforts
of more than 50 technicians, engineers and scientists. Integrating the safety and management approaches of the Chinese and U.S. groups  was a
considerable challenge.  Detailed installation and assembly procedures were devised to endure both the safety of workers 
and equipment.  These procedures were instrumental in  preventing personal injury to any  workers. 
The procedures were not perfect, however,  some incidents described in the previous section could have been prevented by 
improved adherence to the procedure or improved equipment checks.

Lessons learned during the prototype assembly speeded assembly of the first detector pair. As experience in assembling the detector grew,
assembly times became shorter.
The first detector pair took about 14 months to assemble.
The third pair took less than 4 months to assemble.  Installation speed in the Experimental Halls also improved by a factor of two from
the first detectors to the last detectors ending at 2-3 weeks per detector. 

The adopted AD design was able to accommodate  to the differences in dimensions of the as-built AD components. In general,
most physical dimensions were within 0.5\% of each other or better, ensuring that the ADs were functionally identical. 
A significant amount of time and effort was spent in leak checking and quality assurance checks. The first two ADs have operated reliably 
for nearly two years. The latter detectors have also had no significant problems.   
Systematic errors associated with the detector are already  less than the original design goals 
and will be improved further in the coming years.

\acknowledgments

We would like to thank the U.S. and Chinese technicians who made the timely completion of this experiment possible.
This work was supported in part by the DOE Office of Science, High Energy Physics, under contract DE-FG02-95ER40896, the University of Wisconsin, the Alfred P. Sloan Foundation,
 the Research Grants Council of the Hong Kong Special Administrative Region of China (Project Nos. CUHK 1/07C and CUHK3/CRF/10), and the focused investment scheme of CUHK

\end{document}